%% file: main.tex
\documentclass[journal]{vgtc}                     %
\usepackage{balance}

\newcommand{\add}[1]{\textcolor{black}{#1}}

\onlineid{1836}

\vgtccategory{Research}

\vgtcpapertype{Theoretical \& Empirical}

\title{Shape It Up: An Empirically Grounded Approach \\ for Designing Shape Palettes}

\author{%
  \authororcid{Chin Tseng}{0000-0003-2025-0755},
  \authororcid{Arran Zeyu Wang}{0000-0002-7491-7570}, \authororcid{Ghulam Jilani Quadri}{0000-0002-8054-5048}, and 
  \authororcid{Danielle Albers Szafir}{0000-0003-3634-8597}
}

\authorfooter{
  \vspace{-1mm}
  \item
  	Chin Tseng, Arran Zeyu Wang, and Danielle Albers Szafir are with the University of North Carolina at Chapel Hill. E-mail: \{chint, zeyuwang, danielle.szafir\}@cs.unc.edu
  \item
  	Ghulam Jilani Quadri is with the University of Oklahoma and University of North Carolina at Chapel Hill.
  	E-mail: quadri@ou.edu
}

\abstract{
Shape is commonly used to distinguish between categories in multi-class scatterplots.
However, existing guidelines for choosing effective shape palettes rely largely on intuition and do not consider how these needs may change as the number of categories increases. 
Unlike color, shapes can not be represented by a numerical space, making it difficult to propose general guidelines or design heuristics for using shape effectively. 
This paper presents a series of four experiments evaluating the efficiency of 39 shapes across three tasks: relative mean judgment tasks, expert preference, and correlation estimation. 
Our results show that conventional means for reasoning about shapes, such as filled versus unfilled, are insufficient to inform effective palette design. Further, even expert palettes vary significantly in their use of shape and corresponding effectiveness. 
To support effective shape palette design, 
we developed a model based on pairwise relations between shapes in our experiments and the number of shapes required
for a given design. 
We embed this model in a palette design tool to give designers agency over shape selection while incorporating empirical elements of perceptual performance captured in our study.
Our model advances understanding of shape perception in visualization contexts and provides practical design guidelines that can help improve categorical data encodings. 
}


\teaser{
  \centering
  \includegraphics[width=\linewidth, alt={A view of a city with buildings peeking out of the clouds.}]{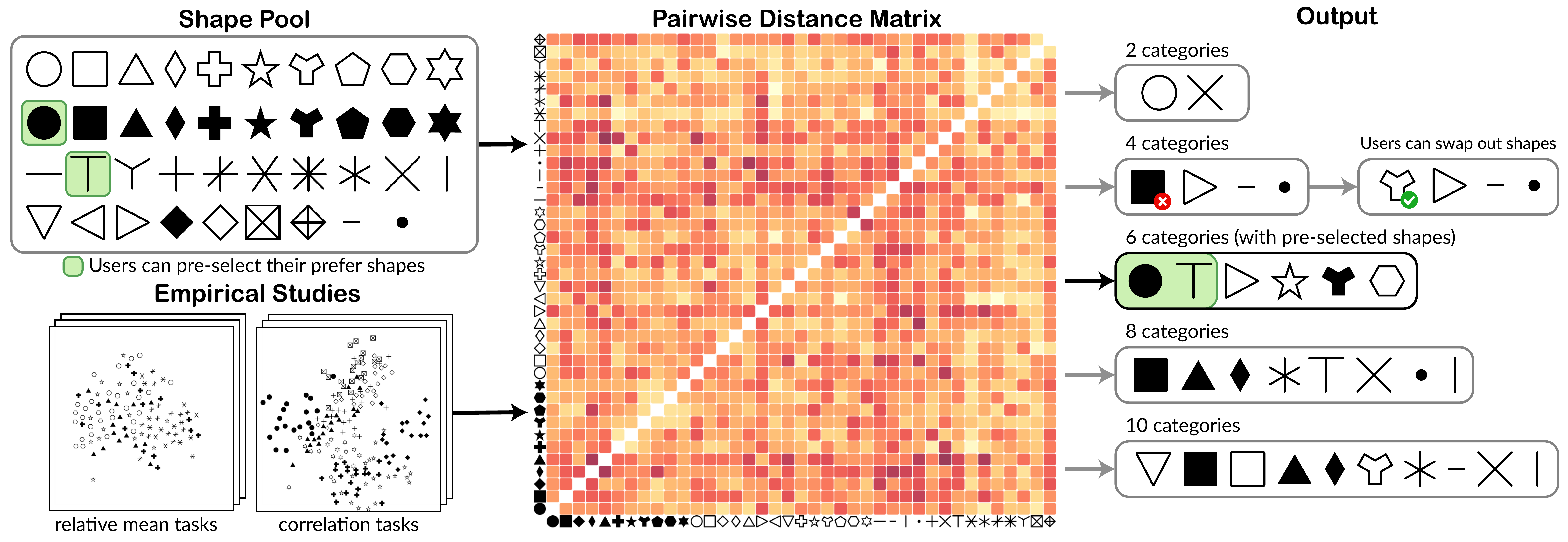}
  \caption{We present a web-based shape recommendation tool based on our empirical studies. 
  Users can input their target category number and preferred shape, and the tool will provide a shape palette based on a pairwise distance model between shapes generated using our experimental results. 
  The output shape palette can also be modified by swapping out certain shapes, which the system will replace using data-driven recommendations. }
  \label{fig:teaser}
}

\graphicspath{{figs/}{figures/}{pictures/}{images/}{./}} %

\usepackage{tabu}                      %
\usepackage{booktabs}                  %
\usepackage{lipsum}                    %
\usepackage{mwe}                       %

\usepackage{mathptmx}                  %

\begin{document}

\maketitle

\section{Introduction}
\input{sec-intro}

\section{Related Work}
\input{sec-related}

\section{Experiment One: Shape Types}
\label{sec-exp1}
\input{sec-exp1}

\section{Experiment Two: Shape Palette}
\label{sec-exp2}
\input{sec-exp2}

\section{Experiment Three: Experts’ Preference}
\label{sec-exp3}
\input{sec-exp3}

\section{Experiment Four: Shape and Correlation}
\label{sec-exp4}
\input{sec-exp4}

\section{Modeling Shape Differences}
\label{sec-tool}

\input{sec-tool}

\section{Discussion}

\input{sec-discussion}

\section{Conclusion}
\input{sec-conclusion}
\acknowledgments{
We thank the reviewers for their insightful comments.
This work was supported by the National Science Foundation under
grant No.2127309 to the Computing Research Association for the
CIFellows project, NSF IIS-2046725, and NSF IIS-1764089.
}

\balance
\bibliographystyle{abbrv-doi-hyperref}
\bibliography{main}

\appendix %

\end{document}

%% file: sec-intro.tex
Multiclass scatterplots allow people to compare patterns between different categories of data. Scatterplots typically use either shape or color to encode these categories. While color is a popular and well-studied method for representing categorical data~\cite{gramazio2016colorgorical,tseng2023evaluating,harrower2003colorbrewer,kim2018assessing}, it can be inaccessible for people with color vision deficiencies \cite{elavsky2022accessible}. 
Many systems instead default to using shape to encode categories in multiclass scatterplots~\cite{demiralp2014learning, ware2009quantitative, burlinson2017open}.
However, relatively little attention has been paid to how to use shapes to encode categorical data, especially with respect to guidelines for selecting sets of shapes that support effective multiclass analysis and for understanding the robustness of shapes as the number of categories grows.

Unlike colors, which can be mapped to continuous numerical color spaces, we do not have an accepted mathematical space for comparing shapes, and building such a model remains an open research question \cite{li2020validated}. Instead, past work in visualization typically characterizes shapes according to their broad features (e.g., open versus closed \cite{burlinson2017open}, filled versus unfilled \cite{smart2019measuring}, perceived order \cite{ware2009quantitative}). Shape descriptors \cite{heider2011local} characterize differences in shapes based on target features. However, we do not yet know the role shape features play in assembling effective palettes. 
Most tools rely on shape palettes preassembled by designers. The composition of these palettes tends to lean more on personal preference and experience rather than empirical evidence. We lack actionable, empirically grounded insight into what dimensions of shape lead to effective shape palettes for visualization. In this paper, we conduct a series of studies examining the effectiveness of shape palettes for encoding categorical data. 

We explore shape palette design from three perspectives: perceptual type (i.e., open, filled, and unfilled \cite{burlinson2017open,smart2019measuring}), expert palettes, and distance modeling \cite{demiralp2014learning}.  
Perceptual type characterizes the general visual structure of a set of shapes \cite{smart2019measuring,burlinson2017open}. For example, Burlinson et al.~\cite{burlinson2017open} found that shapes with a closed contour (e.g., triangles or squares) tended to be more effective than shapes with open contours (e.g., crosses or asterisks) and that mixing shape types reduced performance. However, their studies focused on two-class scatterplots. 
Visualization tools like D3~\cite{6064996}, Tableau~\cite{tableau}, Excel~\cite{msexcel}, and Matlab~\cite{MATLAB} provide predefined shape palettes; however, these palettes are largely built on heuristics and designer preferences rather than on empirically grounded or actionable principles for shape palette composition beyond shapes needing to be ``simple'' and ``discriminable''~\cite{casperson1950visual, deutsch1955theory}.
Demiralp et al.~\cite{demiralp2014learning} instead leverages an empirical approach to modeling shape similarity, characterizing ten shapes according to different similarity measures. However, as with color \cite{stone2006choosing,gramazio2016colorgorical,schloss2018mapping}, shape similarity alone may be insufficient to characterize what shapes compose effective palettes. 
We analyze these perspectives on shape palette effectiveness in a series of scatterplot estimation tasks. We use the results of these experiments to collectively inform a preliminary model to support effective shape palette design.  

\add{We conducted four experiments to analyze how shapes impact people's abilities to distinguish between categories.
These studies first explored how well people compared the relative means of different categories using different shape types (Experiment One) and designer palettes (Experiment Two). The results of these experiments indicated that effective shape palette design is complex and not predicted by any intuitive set of shape features. To more closely examine palette design, we examined subjective palette preferences from 21 visualization experts (Experiment Three) and expanded our objective data to support a computational model for palette design guidance by asking people to compare correlations across categories (Experiment Four).}
Across all studies, we found that shape is complex: no obvious feature or set of features defined an ``effective palette.''
To offer actionable support for palette design, we 
used our data to construct a preliminary model that guides palette generation for up to ten categories. 
We implemented this model in an interactive palette design tool, enabling users to input their preferred shapes and required number of categories, thus providing agency for designers creating effective shape palettes.

\noindent \textbf{Contribution:} We investigated how shape types and designer-crafted shape palettes influence people's perception of multiclass scatterplots across varying category numbers. 
We constructed a model evaluating the perceptual distances for each pair from 39 shapes based on the results from four experiments. 
We implement this model in a web-based palette recommendation \textcolor{blue}{\href{https://shape-it-up-eec5e.web.app/}{tool}} providing insights into effective shape palettes based on empirical studies.

%% file: sec-related.tex
Empirically grounded guidelines support more effective data visualization \cite{elliott2020design,quadri2021constructing}.
Understanding shape palette design for categorical data encoding %
can improve multiclass scatterplot design. However, 
we lack insight into how different shape design choices impact perceptual effectiveness in multiclass scatterplots.
We briefly review related literature on perception studies in scatterplots, categorical perception in visualization, and shape palette design
to ground our 
approach.

\subsection{Graphical Perception in Scatterplots}
Scatterplots are 
both widely studied and often used to encode categorical data~\cite{sarikaya2018scatterplots, quadri2021survey, munzner2014visualization}.
Rensink \cite{rensink2018information} notes that scatterplots serve as a sort of visualization ``fruit fly,'' providing a simple, comprehensible paradigm for studying how a range of design choices influence visualization perception. 
Recent studies have used scatterplots to explore the impact of various visual channels such as 
color~\cite{tseng2023evaluating, tseng2024revisiting, gramazio2016colorgorical}, point size~\cite{hong2021weighted}, shape categories~\cite{burlinson2017open}, and opacity~\cite{micallef2017towards}, as well their combined effects across multiple visual channels~\cite{smart2019measuring, szafir2018modeling, demiralp2014learning,quadri2020modeling,quadri2022automatic,gleicher2013perception} on perceptual performance.
These studies highlight 
\add{how effective visual design decisions can significantly improve perceptual efficiency}
in 
different tasks, including correlation estimation~\cite{rensink2010perception, harrison2014ranking, kay2015beyond}, mean estimation~\cite{tseng2023evaluating,tseng2024revisiting, gleicher2013perception}, cluster identification~\cite{wang2019improving, abbas2019clustme, jeon2023clams}, and other complex data patterns~\cite{quadri2024do, szafir2023visualization, urribarri2017prediction}.

Scatterplots also serve as a primary visualization idiom for generating many actionable design models.  
\add{Several statistical models and visual quality measures 
characterize specific visual patterns in scatterplots to help users conduct different tasks~\cite{wilkinson2005graph, abbas2019clustme,jeon2023clams}. For example, Scagnostics \cite{wilkinson2005graph} summarizes common patterns in scatterplots' visual structure.}
However, these techniques leverage specialized measurements with high learning costs and may not always be grounded in empirical studies of visualization perception~\cite{wang2019improving}. Metrics like ClustMe \cite{abbas2019clustme} use data from empirical studies to generate actionable measures of scatterplot properties grounded in human perception. 
We build on these traditions to understand effective shape palette design, with a focus on palette design in multiclass scatterplots. We draw on approaches for generating empirically grounded models to use our experimental data to provide actionable guidance for effective palettes.

\subsection{Categorical Perception in Visualization}

Categorical 
data represents differences in groups and lacks magnitude or order (e.g., ``cats'' and ``dogs'') \cite{munzner2014visualization}. 
Many visualization idioms, such as bar charts,
often rely on color or texture differentiation to communicate categories. Scatterplot marks can use both color and shape to enhance categorical distinction
\cite{ali2013effect, quadri2021survey, quadri2024do}.
In scatterplots, the specific choice of shape and color to represent different categories can affect 
people's abilities to reason across categories~\cite{tseng2023evaluating, smart2019measuring, burlinson2017open, ware2012information, demiralp2014learning}.
For example, Burlinson et al. \cite{burlinson2017open} found that using pairs of closed shapes supported more effective analysis in two-class scatterplots compared to pairs of open shapes or mixed closed-open pairs. 
Understanding categorical encoding perception helps designers understand and predict how viewers discern and process visual categories in estimating statistical quantities but is 
still underexplored 
in graphical perception~\cite{goldstone2010categorical,szafir2023visualization}.

Factors such as the number of categories may directly influence our abilities to reason across categorical data~\cite{tseng2023evaluating}.
Most existing design guidance for encoding categorical data describes how to use color to encode categories. 
Our abilities to reason across categorical colors is closely tied to the discriminability of the color encoding~\cite{tseng2023evaluating, haroz2012capacity}, with more distinct colors typically providing greater discriminability.
Distinct color hues, rather than colors of different lightness or saturation, 
allow for more accurate visual mean comparison~\cite{tseng2024revisiting, tseng2023evaluating}. Using distinctly nameable colors may also enhance our abilities to distinguish different color categories \cite{heer2012color}.
However, factors beyond color difference also influence the effectiveness of categorical color palettes. 
For example, colors in categorical data are more effective when they align with the semantic concepts associated with  categories \cite{schloss2018mapping,mukherjee2021context,lin2013selecting,setlur2016linguistic}.

Similarly, shapes can encode categorical data. Shapes often 
are differentiated across a range of features~\cite{ware2012information, demiralp2014learning, deutsch1955theory}.
While general guidelines suggest that shapes should be selected based on their distinctiveness and the ease with which they can be effectively processed~\cite{demiralp2014learning}, we lack systematic ways for reasoning about these factors in practice.
Past studies investigated coarse-grained distinctions between shapes in categorical data encoding \cite{burlinson2017open} and introduced methods for generating shapes based on fractal geometries \cite{ebert2000procedural,shaw1998data} or generative polygons \cite{brehmer2021generative}. However, these approaches rely on broad intuitions about potential features in shape that may matter for categorical encodings. The ways people process shapes are complicated and still not fully understood \cite{loffler2008perception,pasupathy2018visual,baker2018abstract}. A range of potential visual features may influence shape perception \cite{dai2022visual,li2020validated}, and the role of these features in categorical encoding specifically is not well understood. In this paper, we aim to understand how shapes impact categorical perception in multiclass scatterplots. We approach this problem using existing research and practice in visualization as a 
\add{basis} for understanding palette design. 

\subsection{Shape Palettes Design and Perception}

While color palettes have been extensively studied, shape palettes pose unique challenges in part due to the lack of a grounded numeric space for defining and reasoning about relationships between shapes. 
The perceptual effectiveness of shapes in visualization is influenced by various factors, including visual channels like mark size and color, the context of the visualization, and the inherent properties of the shapes themselves~\cite{demiralp2014learning, smart2019measuring, szafir2018modeling, ware2012information}.
Designing effective shape palettes is challenging because it requires selecting shapes that are both distinct from one another and capable of effectively and robustly conveying data in categorical and statistical quantities. However, shapes may interact in different ways \cite{burlinson2017open}, and factors that influence color palettes, like semantics or aesthetics, likely play a role in shape palette effectiveness but are not well understood in this context.

Common shapes in visualization palettes include circles, squares, and triangles, though visualizations may use more complex or procedurally generated shapes
such as superquadric glyphs~\cite{shaw1998data} and 
diatoms~\cite{brehmer2021generative}.
While palettes like Q-Tons \cite{ware2009quantitative} are designed to communicate order, most shape palettes simply aim to present a set of distinct shapes. 
Past research typically categorizes shapes into three types: \textit{filled, unfilled}, and \textit{open} shapes~\cite{smart2019measuring, burlinson2017open, ware2009quantitative, ware2012information} (see \autoref{fig:shape-types} for examples). While geometric shape descriptors can characterize specific features of shapes \cite{heider2011local}, these descriptors often focus on more complex geometries than the simple and often nameable glyphs used in most shape palettes.

Designing a shape palette requires careful consideration of the visualization’s context and data.
Many visualization tools offer predefined designer-crafted shape palettes that are hand-tuned for clarity and discriminability.
Similar to color palettes, the goal of these palettes is typically to create a set of shapes that are easily distinguishable yet harmonious.
Strategies include using a mix of geometric and organic shapes catering to different data types or employing a consistent shape style that aligns with the visualization’s overall aesthetic and data's natural hierarchies~\cite{krzywinski2013plotting}.
Most visualization tools adopt their own palette design (see \autoref{fig:shape-palettes} for examples). 
For example, 
Excel~\cite{msexcel} has a palette of nine filled and open shapes. 
Tableau~\cite{tableau}, R~\cite{R}, and Matlab~\cite{MATLAB} have ten unfilled and open shapes, but only half of those shapes are shared across the three palettes. 
These tools often follow designer-suggested practices in shape palette design, such as offering a variety of shapes with distinct silhouettes and avoiding overly complex designs that can be difficult to discern at a glance~\cite{brath2010multiple}.
However, these heuristics only provide rough guidelines with significant subjectivity in their implementation, as seen in the diversity of designer palettes. We instead aim to better understand how we can help people more actionably create effective shape palettes for data visualization. We do so using a series of experiments measuring how well different shape palettes support categorical data encoding.

%% file: sec-exp1.tex
\begin{figure*}[htbp] 
\centering
\vspace{-2em}
\includegraphics[width=1\textwidth]{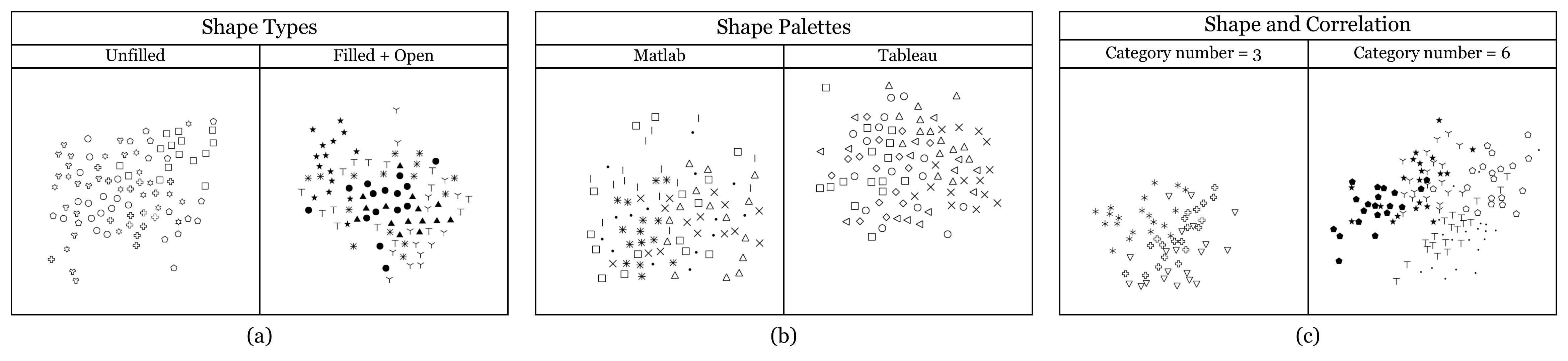} 
\vspace{-2em}
\caption{(a) Two examples of stimuli used in Experiment 1, both with six categories encoded with single-type (unfilled) and two-type (filled + open). (b) Two scatterplots used in Experiment 2, encoded with different shape palettes from Matlab and Tableau, both with six categories. (c) Two scatterplots with different category numbers (3 and 6) used in Experiment 4 for measuring pairwise distances. Both (a) and (b) employed relative mean judgment tasks while (c) applied correlation judgment tasks. } 
\vspace{-1em}
\label{fig:stimuli}
\end{figure*}

Our first study analyzed how the choice of shape types (filled, unfilled, and open \cite{burlinson2017open,smart2019measuring}) used to differentiate categories impacts people's abilities to analyze multiclass scatterplots. We performed a crowdsourced study measuring how well people were able to compare category means over varying category numbers ($N=2$--$10$).
We hypothesized that:

\noindent\textbf{H$_{1}$:} \textbf{(a) choices of shape types would impact performance,} and \textbf{(b) palettes using shapes from multiple types would outperform palettes using shapes from a single type.} Shapes can be categorized by their features, and different types of shapes can influence how people perceive data. While Burlinson et al. \cite{burlinson2017open} found that mixing shape types may impair performance for two-shape palettes, we expect people may benefit from enlarging the discriminability between shapes, which likely requires integrating multiple shape types when analyzing larger numbers of categories. 

\subsection{Experiment Design}
\subsubsection{Task}
We employed a relative mean judgment task 
as used in previous studies of multiclass scatterplot design~\cite{gleicher2013perception,hong2021weighted,kramer2017visual,tseng2023evaluating}. 
This task asked participants to estimate the category with the highest average y-value.  We selected this task because it requires participants to locate data points across various categories and subsequently calculate statistical values across all points within each category. Confusion between points from different categories is reflected in participants’ responses, as this task is subjected to both overinclusion (i.e., including points that are not from certain categories) and underinclusion (i.e., failing to include points from certain categories).

\begin{figure}[htbp] 
\centering
\vspace{-0.2em}
\includegraphics[width=0.5\textwidth]{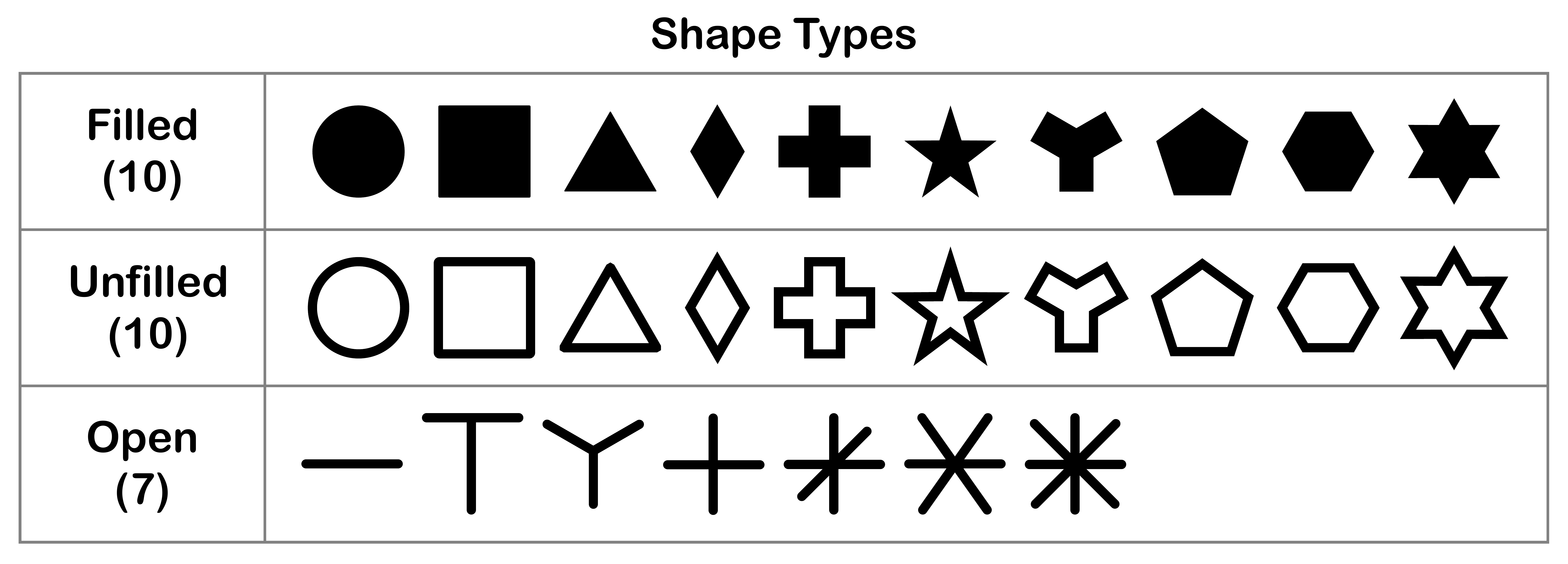} 
\vspace{-1em}
\caption{We collected shapes from multiple sources and categorized them into three shape types: filled, unfilled, and open. Both filled and unfilled have 10 shapes and open type has 7 shapes.} 
\vspace{-1.5em}
\label{fig:shape-types}
\end{figure}

\subsubsection{Stimuli Generation}
Participants estimated category means using a series of black-and-white scatterplots using shape to differentiate categories. 
As shown in \autoref{fig:stimuli} (a), we generated each scatterplot as a 400x400 pixel graph using D3. Each
scatterplot was rendered to a white background and two orthogonal
black axes with 13 unlabeled ticks. 

Categories were encoded using a subset of 27 shapes
collected from
previous studies~\cite{smart2019measuring, burlinson2017open, ware2009quantitative}, popular programming libraries (Matplotlib~\cite{Hunter:2007}, Matlab~\cite{MATLAB}, R~\cite{R}, D3~\cite{6064996}, and Plotly~\cite{plotly}), and commercial visualization tools (Tableau~\cite{tableau} and Excel~\cite{msexcel}). We assembled this corpus by collating all predefined shapes used to encode data in these 10 sources. We removed duplicates, 
identified common shapes (i.e., those occurring in more than one palette), 
and then excluded any shapes that were rotations of other shapes in the set. 
Finally, we categorized these shapes into three groups consistent with prior work~\cite{burlinson2017open,smart2019measuring}: filled shapes, unfilled shapes, and open shapes. Our final shape set is shown in \autoref{fig:shape-types}.

Shapes were rendered in a 6$\times$6-pixel window, with window size chosen in piloting and reflecting prior studies~\cite{tseng2023evaluating}. 
For shapes that were intentionally smaller on the x- or y-dimension, like the diamond from D3's default palette, or smaller than other shapes in our palette, like the dot from Matlab or the half line from Excel, %
we preserved their aspect ratio and size relative to the original shape. 

Each dataset used in the study contained between $N = 2 -10$ categories, with each category mapped to a unique shape. Datasets drew from 2D Gaussian normal distribution \add{ranging in [0, 1]} with random \add{x-axis and y-axis} means ranging between [0.1, 0.9]
with overlapping points jittered \add{to avoid confounds from spatial overlap}. Each category contained 20 datapoints and the mean difference between categories with the highest y-mean and second-highest y-mean was in [0.2, 0.25], corresponding to moderate hardness in prior studies \cite{tseng2023evaluating}.  

We compare performance both between three shape types individually and across combinations of types. 
We created three groups---single-type (filled, unfilled, open), two-type (filled + unfilled, filled + open, unfilled + open), and three-type (filled + unfilled + open)---that represent all combinations of categories. 
As testing all possible combinations of 27 shapes and 9 category numbers is intractable, we generated 
ten shape combinations for each type group and category number (N=2--10). The shape sets were randomly selected with replacement based on 
the type group and category numbers.
The number of shapes was equally distributed from each type and randomly picked within a type. This generation method created 750 total shape combinations.

\subsubsection{Procedure}
Our experiment consisted of three phases: (1) informed consent, (2) task description and tutorial, and (3) formal study. Participants first provided informed consent under our IRB protocol and then asked to provide demographics. Participants were then introduced to the mean judgment task description and led to the tutorial section, where they completed three tutorial questions, asking them to complete the target task with 2-3 categories with filled circle, unfilled circle, and open asterisk. They were required to successfully answer all the tutorial questions before proceeding to reduce possible ambiguities 
in task understanding. 

During the formal study, participants completed our target task (``identify the category with the highest average y-value'') for 53 stimuli presented sequentially (50 formal
trials and three engagement checks). To balance the number of both shape types and category numbers for each participant, we used a stratified sampling of our pre-generated shape combinations (shape types $\times$ category number) to create 15 task groups, with each group containing at least five tasks per category number and seven tasks per shape type. One group was randomly assigned to each participant at the beginning of the study, with stimuli within each group presented in a random sequential order. Participants had 20 seconds to respond to each stimuli, after which time the answer was marked as incorrect and the study advanced to the next trial. 

To ensure valid participation, we employed three engagement checks. These engagement checks were stimuli with two or three classes that had at least a 0.35 difference in their means. We randomly placed the engagement checks throughout the 50 formal trials.

\subsubsection{Participants}
We recruited 165 participants on Amazon Mechanical Turk (MTurk)
with at least a 95\% approval rating and located within the US and Canada. 15 participants who failed more than one engagement check were excluded. We analyzed data from the remaining 150 participants (115 male, 35 female; 24--62 years of age), \add{with 10 participants in each task group.} All participants reported normal or corrected to normal vision. Our experiment took 10 minutes on average, \add{and each participant was compensated \$1.60 for their time.}

\subsubsection{Analysis}
We used accuracy as our primary dependent measure. To compare the performance between different shape types (filled, unfilled, open) and using shapes from different types (single-type, two-type, three-type), \add{we analyzed the resulting data using a generalized linear model (GLM) with shape type groups and category numbers as independent variables. We used a GLM as the open types have a smaller number of shapes than others, and this model is better suited to the corresponding differences in the response distribution. }
\autoref{tab:exp1} summarizes our results. The anonymized data and results for our study can be found on \href{https://osf.io/5k47c/?view_only=52e6b52f69b84ceab8c8c1b897083fc3}{OSF.}

\begin{table}[htbp] 
\centering
\vspace{-0.5em}
\caption{\add{GLM} results for category number and shape types. Significant effects are indicated by \textbf{bold} text.}
\vspace{-0.5em}
\includegraphics[width=\linewidth]{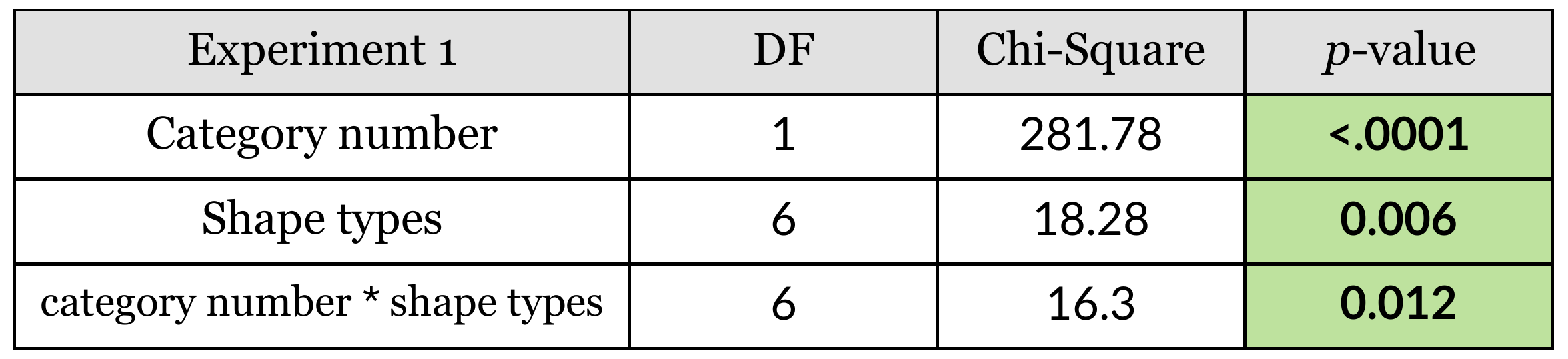} 
\vspace{-0.5em}
\label{tab:exp1}
\end{table}

\subsection{Results}

\label{sec:shapetype}
\begin{figure*}[htbp] 
\centering
\includegraphics[width=1\textwidth]{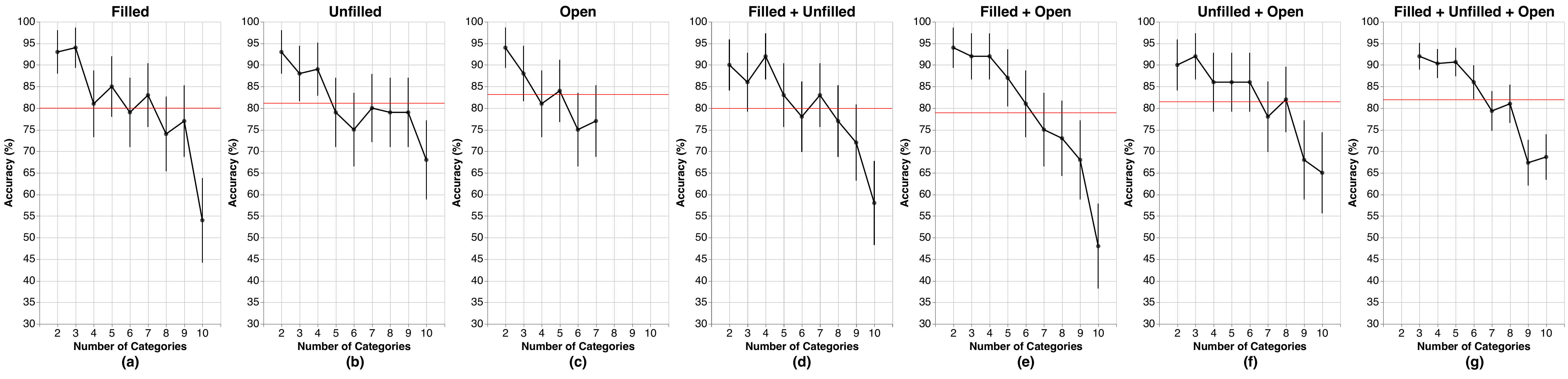} 
\caption{The average accuracy of mean judgment task separated by different shape types and type group combinations in Experiment 1. \add{Overall group means are indicated in red, with category number on the x-axis.}
We used the scale from 30\%-100\% as the slope of the graph \add{(i.e., how robust each group is to increasing numbers of categories)} is the key signal in the data. 
Error bars represent 95\% confidence intervals. \add{Note that open shapes only supported 2--7 categories.}
} 
\vspace{-1.5em}
\label{fig:acc-types}
\end{figure*}

Our analysis revealed a significant effect of shape type on judgment performance 
\add{($\chi^2 (6, N=7500) = 18.28, p = 0.006$).}
\autoref{fig:acc-types} illustrates the overall accuracy results from Experiment 1 per shape type.
Open shapes achieved the best performance (avg. 83.1\%, see \autoref{fig:acc-types}(c)); however, 
 this is in part due to only having 7 open shapes \add{(see below for further analysis).}
Combining all three shape types together resulted in the next-highest accuracy
at 82.0\% (95\% CI: [80.4\%, 83.5\%]) (\autoref{fig:acc-types}(g)).
The mix of filled+open shapes achieved the worst performance (avg. 78.9\%, CI: [76.2\%, 81/6\%], see \autoref{fig:acc-types}(e)).
However, this poor performance was largely a function of performance for larger category numbers: for fewer than six categories, filled+open shapes achieved the best performance (91.3\%, CI: [88.5\%, 94.1\%]), while unfilled shapes were best for larger category numbers (76.5\%, CI: [72.3\%, 80.7\%]). The contrast between filled+open for small and large category numbers suggests that such palettes are effective in some cases, but not robust as the number of categories increases.

As open shapes supported only up to seven categories based on our sample of shapes from commercial tools and combining all three types requires at least three categories, we also 
conducted an exploratory analysis of the data containing between three and seven categories.
Combining all three shape types achieved the highest accuracy (avg. 87.7\%, CI: [86.0\%, 89.3\%]), while open shapes achieved the worst performance (avg. 81\%, CI: [77.6\%, 84.4\%]). 
Using a combination of all three shape types outperformed both open (accuracy difference: 6.7\%, CI: [1.3\%, 12.1\%]) and unfilled shapes (accuracy difference: 5.5\%, CI: [0.05\%, 10.9\%]).

We also found a significant interaction between category number and shape type group 
\add{($\chi^2 (6, N=7500) = 16.3, p = 0.012$).}
As shown in \autoref{fig:acc-types}, filled, unfilled, and open shapes are less robust for fewer categories, while filled+unfilled and filled+open shapes are less robust for larger category numbers.

Our results support \textbf{H$_{1a}$}: we found \add{\textbf{shape types significantly impact performance.}} However, we failed to find significant performance differences between one-type conditions, which means no single shape type (filled, unfilled, or open) significantly outperformed the other.
Our results partially supported \textbf{H$_{1b}$}, \add{\textbf{encodings combining three shape types outperformed both unfilled shapes alone and open shapes alone.}} However, we found little other evidence that using shapes from multiple types outperforms using shapes from fewer shape types (e.g. using two-type compared to single-type).

%% file: sec-exp2.tex
While Experiment 1 found that shape type features can, in part, predict encoding effectiveness, shape type alone only provided limited insight into shape encoding design. Professional designers have created a range of shape palettes that provide sets of shapes used to encode categorical data. However, we lack empirical guidance for effective shape palette design. 
Experiment 2 focused on how existing shape palettes from popular visualization tools influence the performance of relative mean judgment tasks in multiclass scatterplots to understand designer intuition about palette composition. We hypothesized that:

\noindent\textbf{H$_{2}$:} 
\add{\textbf{choices of shape palettes would impact participants' mean estimation accuracy.}}
Different existing shape palettes contain a range of shapes without necessarily following any unified heuristics or guidance. 
Some palettes use orientation to distinguish shapes (e.g., a triangle pointing up and a triangle pointing down). 
Some palettes combine different shape types (e.g., unfilled + open), while some only use single shape type. We expect different design strategies for constructing shape palettes will impact people's 
abilities to distinguish between categories.

\begin{figure}[t] 
\centering
\includegraphics[width=0.5\textwidth]{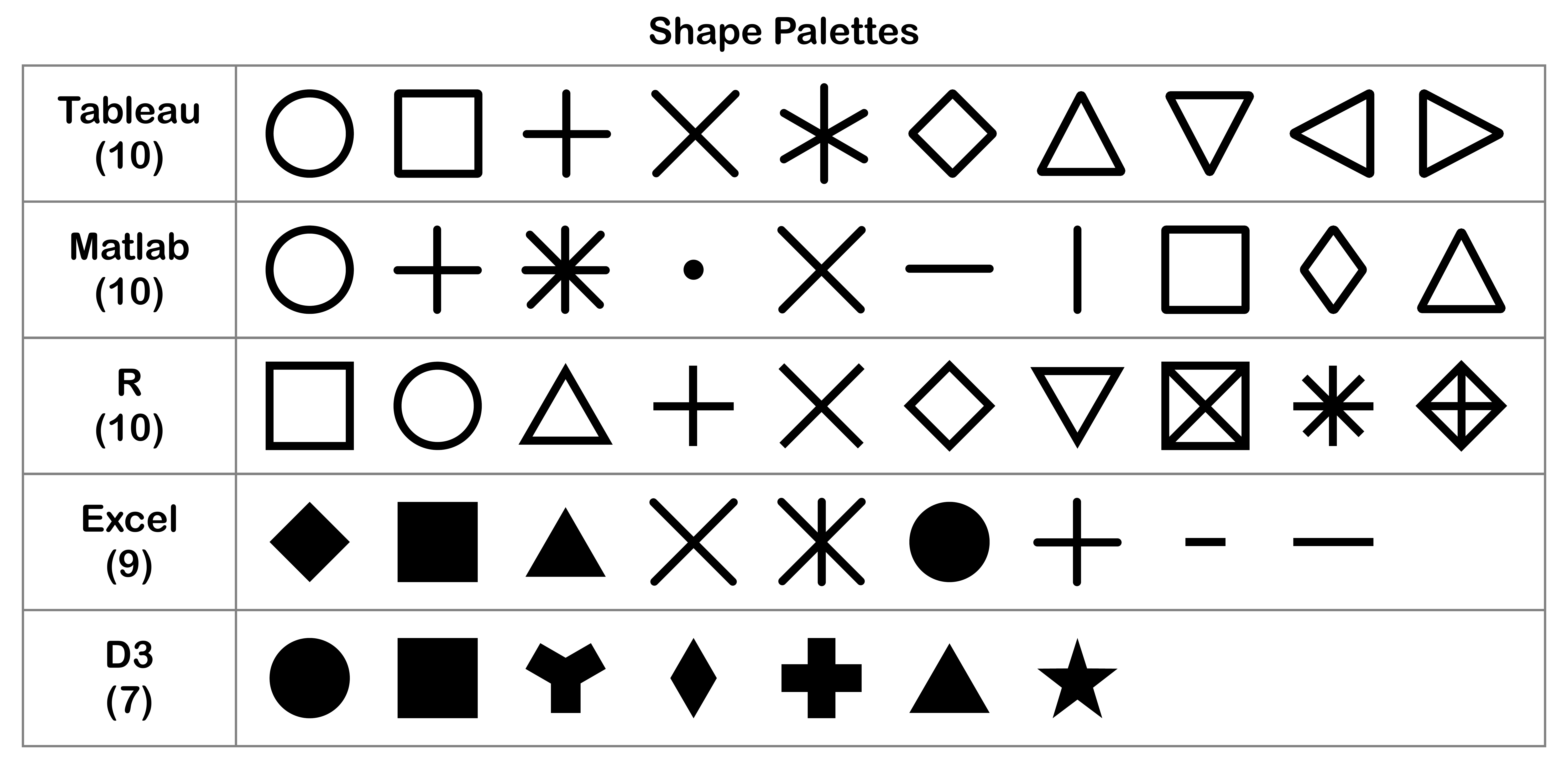} 
\vspace{-1.5em}
\caption{We selected five shape palettes from common visualization tools, including Tableau, Matlab, R, Excel, and D3. Tableau, Matlab, and R have 10 shapes, Excel has 9 shapes and D3 has 7 shapes.}
\vspace{-1.5em}
\label{fig:shape-palettes}
\end{figure}

\subsection{Experiment Design}
\subsubsection{Task \& Stimuli Generation}
We employed the same task, point distribution, point numbers, and difficulty levels as Experiment 1. As shown in \autoref{fig:shape-palettes}, we selected five shape palettes from popular visualization tools: D3~\cite{6064996}, Tableau~\cite{tableau}, Excel~\cite{msexcel}, R~\cite{R}, and Matlab~\cite{MATLAB}. Tableau, Matlab, and R have 10 shapes in their palettes, Excel has 9 shapes, and D3 has 7 shapes. Example stimuli are shown in \autoref{fig:stimuli}(b). 

Some tools define a fixed order of shapes when users apply the shape encodings, but some only provide shapes and require users to 
select their target shapes manually. 
Given the inconsistency in these practices, we assigned shapes to categories at random, with each stimulus drawing $N$ shapes at random, where $N \in [2, 10]$. 
We generated ten random shape combination sets for each shape palette $\times$ category number, resulting in 410 sets in total. We divided the sets into 8 task groups with 52 sets each. Each task group consists of at least five tasks for each category number and seven tasks for each shape palette.

\subsubsection{Procedure \& Participants}
We followed the same general procedure as Experiment 1. Each participant completed 55 trials (52 formal trials and three engagement checks) again presented serially and in a random order. 
89 participants were recruited using the same constraints as Experiment 1. Nine participants who did not pass the engagement checks were excluded, resulting in data from 80 participants (63 male, 17 female; 24--62 years of age) \add{with 10 participants in each task group.} \add{Our experiment took 10 minutes on average, and each participant was compensated \$1.60 for their time.}

\begin{table}[htbp] 
\vspace{-0.5em}
\centering
\caption{\add{GLM} results for category number and shape palettes. Significant effects are indicated by \textbf{bold} text.}
\vspace{-0.5em}
\includegraphics[width=\linewidth]{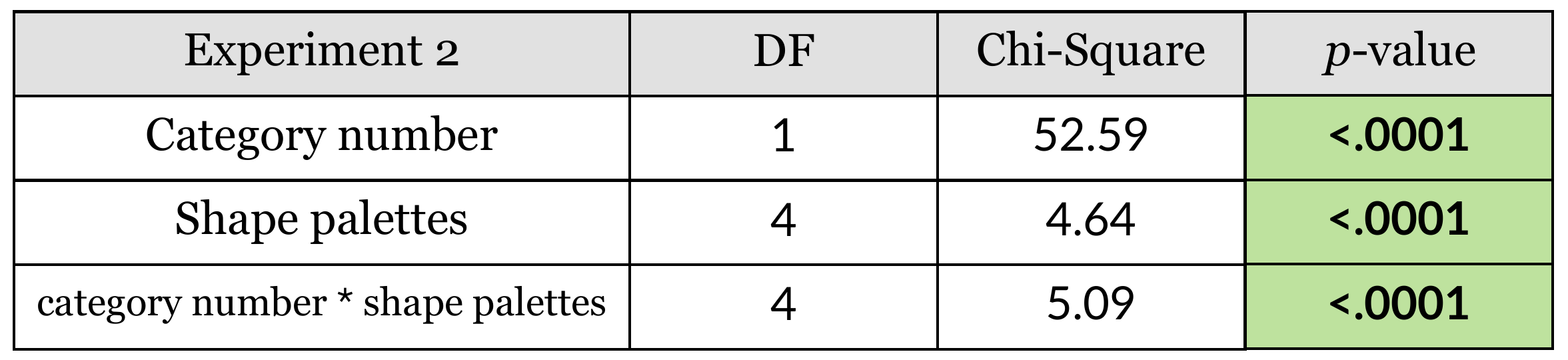} 
\label{tab:exp2}
\vspace{-2em}
\end{table}

\subsection{Results}
\label{sec:shapepalettes}

We again compared response accuracy using a 
\add{generalized linear model (GLM), as partial palettes support a lower number of shapes, 
similar to Experiment 1. }
\autoref{tab:exp2} summarizes our results. 

\begin{figure*}[htbp] 
\centering
\vspace{-2em}
\includegraphics[width=\textwidth]{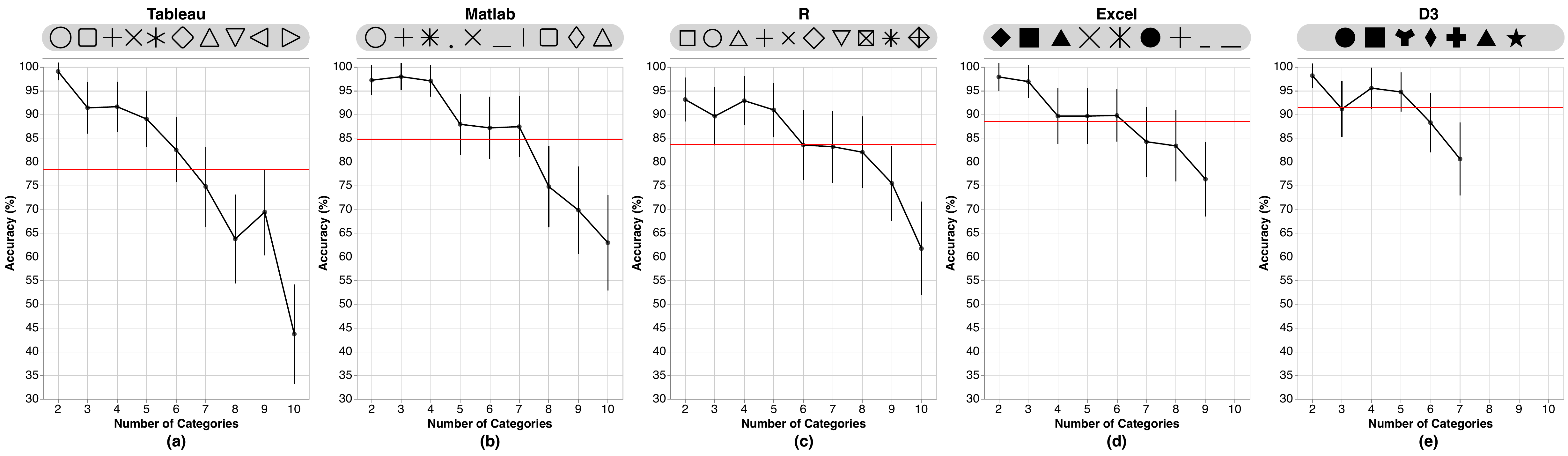} 
\vspace{-1.5em}
\caption{The average mean judgment task accuracy of different shape palettes (top) from professional tools in Experiment 2, with category number on the x-axis. Each chart presents the average accuracy broken down by category numbers (aggregate accuracy is presented in red). The accuracy drops as the category number increases. The average accuracy ranges from 78\% to 92\%. 
Error bars represent 95\% confidence intervals.
\add{Note that Excel (d) and D3 (e) have 9 and 7 shapes respectively.}
} 
\vspace{-1.5em}
\label{fig:acc-palettes}
\end{figure*}

As in Experiment 1, response accuracy was significantly lower as the number of categories increased 
\add{($\chi^2 (1, N=4160) = 52.59, p < .0001$).}
Our analysis also showed a significant effect of shape palettes on mean judgment accuracy 
\add{($\chi^2 (4, N=4160) = 4.64, p < .0001$).}
\autoref{fig:acc-palettes} illustrates the overall results.
The palettes from D3 (avg. 91.39\%, CI:[89.15\%, 93.63\%], \autoref{fig:acc-palettes}(e)), Excel (avg. 88.23\%, CI: [86.03\%, 90.43\%], \autoref{fig:acc-palettes}(d)), and Matlab (avg. 85.06\%, CI: [82.71\% 97.4\%],  \autoref{fig:acc-palettes}(b)) outperformed the other two palettes.  

However, similar to 
open shapes in \autoref{sec:shapetype}, D3 only has 7 shapes which may 
artificially increase its performance.
\add{We further evaluated the accuracy of only trials with 2--7 categories, and found} Matlab, which leverages open and unfilled shapes, achieved the highest average accuracy rate of 92.41\% (CI: [90.30\%, 94.52\%]).
Further, we found accuracy still varied significantly across palettes, from around 75\% to higher than 90\%, even though the palettes are all expert-crafted.

We also found a significant interaction between palette type and category number 
\add{($\chi^2 (4, N=4160) = 5.09, p < .0001$). }
D3 and Excel tended to remain robust across categories; however, they were also the only palettes that did not have ten shapes. Excel's mix of filled and open shapes mimicked the robust performance of that combination for smaller numbers of categories in Experiment 1; however, unlike Experiment 1, the palette remained relatively robust through nine categories. 

The results support \textbf{H$_{2}$}: we found \add{\textbf{different shape palettes, even provided in professional tools, still perform significantly differently in comparing means.}}
We also found no predictable patterns in palette effectiveness based on their constituent shape features. For example, Matlab and Tableau both use a mix of open and unfilled shapes, yet Matlab's palette exhibited significantly higher performance overall.

%% file: sec-exp3.tex
Based on Experiments 1 and 2, 
shape selection has a significant impact on categorical perception.
However, 
we found no discernible pattern in terms of shape features as to which shape combinations are likely to be more accurate or more robust to increasing numbers of categories. Even professional palettes had high variability in performance and robustness. 
To better understand potential patterns or implicit heuristics in palette design, we conducted an exploratory study that asked 21 visualization experts to design ten-shape palettes and analyzed patterns in their responses.
We hypothesized that: \textbf{H$_{3}$:} 
\add{\textbf{experts tend to exhibit consistent patterns in designing color palettes.}} We lack concrete, actionable guidance for shape palette design, as evidenced by the diversity of shapes in the palettes in Experiment 2. However, 
we also see consistent use of some shapes in most tools. For example, all Experiment 2 palettes contained a square, circle, cross, and triangle (either filled or unfilled). Looking at larger patterns in expert design may help to elucidate basic principles or heuristics.

\subsection{Experiment Design}

\subsubsection{Task \& Stimuli Generation}
Participants were asked to select ten shapes to form a palette for encoding ten categories in a scatterplot using a web interface. 
Available shapes consisted of the shapes from both Experiments 1 and 2, shown in the \autoref{fig:teaser} shape pool.
Shapes were displayed in random order and rendered in black on a white background. Participants clicked on the desired shapes, which were then outlined, and selected shapes were displayed in a line at the top of the webpage to preview the palette.

\subsubsection{Procedure \& Participants}
21 visualization experts 
voluntarily participated in our study in-person during a workshop at a visualization conference. We collected no data other than a selection of 10 shapes. Our experiment was deployed on a website. Participants submitted their responses anonymously. 

\subsection{Results}

\begin{figure}[tbp] 
\centering
\includegraphics[width=0.48\textwidth]{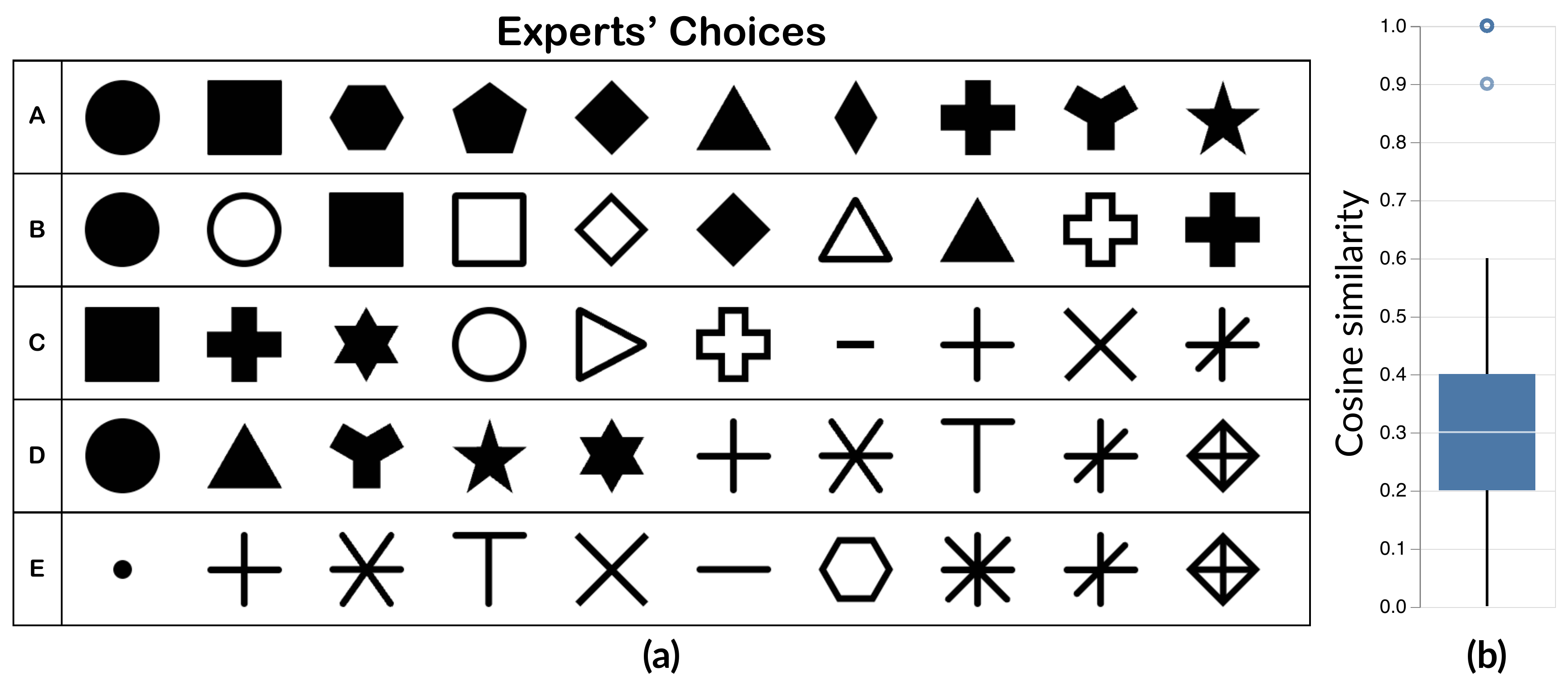} 
\vspace{-0.5em}
\caption{(a) Five instances of expert-chosen shape palettes. (b) Pairwise cosine similarity between 21 experts. Some prefer using single-type (A selected all filled shapes. E selected mostly open shapes), while some prefer multiple types (B, C, and D selected filled + unfilled, filled + unfilled + open, and filled + open respectively). For all the experts' choices, please refer \href{https://select-shapes.web.app/result.html}{here.}
} 
\vspace{-1em}
\label{fig:acc-expert}
\end{figure}

We visually examined differences between expert choices in Experiment 3 when designing their own shape palettes.
\autoref{fig:acc-expert}(a) illustrates five examples of expert-crafted shape palettes embodying different selection strategies.
All expert-chosen palettes are available \href{https://select-shapes.web.app/result.html}{here.}

In this analysis, we did not find any noteable patterns in similarity in shapes among expert-chosen palettes.
Some experts focused on one specific shape type: for example, Expert A (i.e., A in \autoref{fig:acc-expert}(a)) chose all filled shapes; Expert E mostly chose open shapes; and Expert C chose an almost uniform distribution of filled, unfilled, and open shapes.
Other experts 
preferred shapes with similar geometric features. For example, Experts B, C, and D chose shapes that have similar skeletons across different shape types, reminiscent of Q-ton sequences \cite{ware2009quantitative}.

We performed pairwise cosine similarity on 21 selection sequences (\autoref{fig:acc-expert}(b)), finding an average similarity of 0.35 ($\sigma=0.28$). 
The diversity of these results indicates that shape is a sufficiently complex visual channel that \add{\textbf{even visualization experts do not display consistent strategies in shape palette design}}, failing to support \textbf{H$_{3}$}. 

%% file: sec-exp4.tex
Given the lack of consistent palette design guidelines, performance, and practices and based on results from past work and Experiments 1--3, we analyzed how different combinations of shapes
influence people's abilities to reason about correlations within scatterplots to better understand overall palette performance. 
We hypothesized that:
\textbf{H$_{4}$:} \textbf{
The relationships between shapes will influence people's abilities to compare correlations.
} Since people need to identify a category by shape and compare statistics to categories of another shape, we expect that the perceptual distance between shapes will impact people's performance, as seen in the robustness of three-type palettes in Experiment 1.
While we lack a specific model for perceptual difference in shape encodings, Demiralp et al. \cite{demiralp2014learning} and Ware \cite{ware2009quantitative} both demonstrate that shapes can be ``ranked'' by similarity. 
More similar shapes may cause elements from one category to be computed as part of a second category. We draw on the pairwise approach of Demiralp et al. \cite{demiralp2014learning} to model relationships between shapes and use a second task---correlation---to complement the data from Experiments 1 and 2.

\subsection{Experiment Design}
\subsubsection{Task}
We asked people to compare the correlation of different categories, similar to previous studies~\cite{harrison2014ranking, kay2015beyond, rensink2010perception}. While this task likely uses similar perceptual mechanisms to averaging \cite{szafir2016four}, it reflects a second common task people use scatterplots for where estimates are affected by people's abilities to distinguish different categories. 
Participants were asked to estimate which category was the most correlated. 

\subsubsection{Stimuli Generation}
The scatterplots, shape sizes, and point numbers were generated using the same approaches as Experiments 1 and 2, as shown in \autoref{fig:stimuli}(c). We drew shape palettes from the set of 39 shapes in Experiment 3.

Point locations were generated by the random multivariate method from NumPy~\cite{2020NumPy-Array}, which takes mean, covariance, and point number and returns random samples from a multivariate normal distribution. We \add{start with} randomly chosen \add{x and y means from the range [0.1, 0.9], 0.95 covariance for target category and 0.6 covariance for the rest categories.}
 The target category that has the highest Pearson correlation coefficient ranges from 0.8 to 0.95, and the second-highest category has at least a 0.2 correlation difference from the target category. 
\add{We jittered points to avoid overlapping and resampled points until the correlation coefficient values satisfied the criteria.}
While previous studies~\cite{kay2015beyond, harrison2014ranking, rensink2010perception} indicate the just-noticeable difference (JND) for $r=0.8$ ranges from 0.05-0.15, these studies focus on single class and pairwise comparison only. In piloting, we found that greater category numbers significantly reduced performance at these thresholds, with $\Delta r=0.2$ offering approximately 75\% mean performance. 

To understand relationships between shapes, we focus on the performance difference between pairs of shapes, forming a proxy distance metric.  
To generate a pairwise distance matrix with a comparable sampling distribution, we applied a progressive selection strategy for generating the tested shape combinations. We started by randomly picking ten shape combinations for 2 categories. For each set of progressively larger categories, we sorted the shape pairs by the number of entries that test that pair so far
and construct candidate combinations that create the most entries for lower-ranking pairs. We randomly select ten combinations from these candidates. We then increment the number of categories and repeat the process.  
To evaluate our selection method, we generated one shape combination dataset with using random selection and compared it with one with our progressive selection strategy. The number of pairs for the random selection strategy ranged from 7 to 35 ($\sigma=4.3$), while our approach ranged from 18 to 27 ($\sigma=1.7$), suggesting the progressive method provides a more balanced sampling.

We generated 810 shape combination sets using this strategy (90 shape combinations $\times$ 9 category numbers). We divided these sets into 15 task groups, with each category number equally distributed (6 tasks $\times$ 9 category numbers = 54 tasks per group).

\begin{figure*}[t] 
\centering
\vspace{-2em}
\includegraphics[width=1\textwidth]{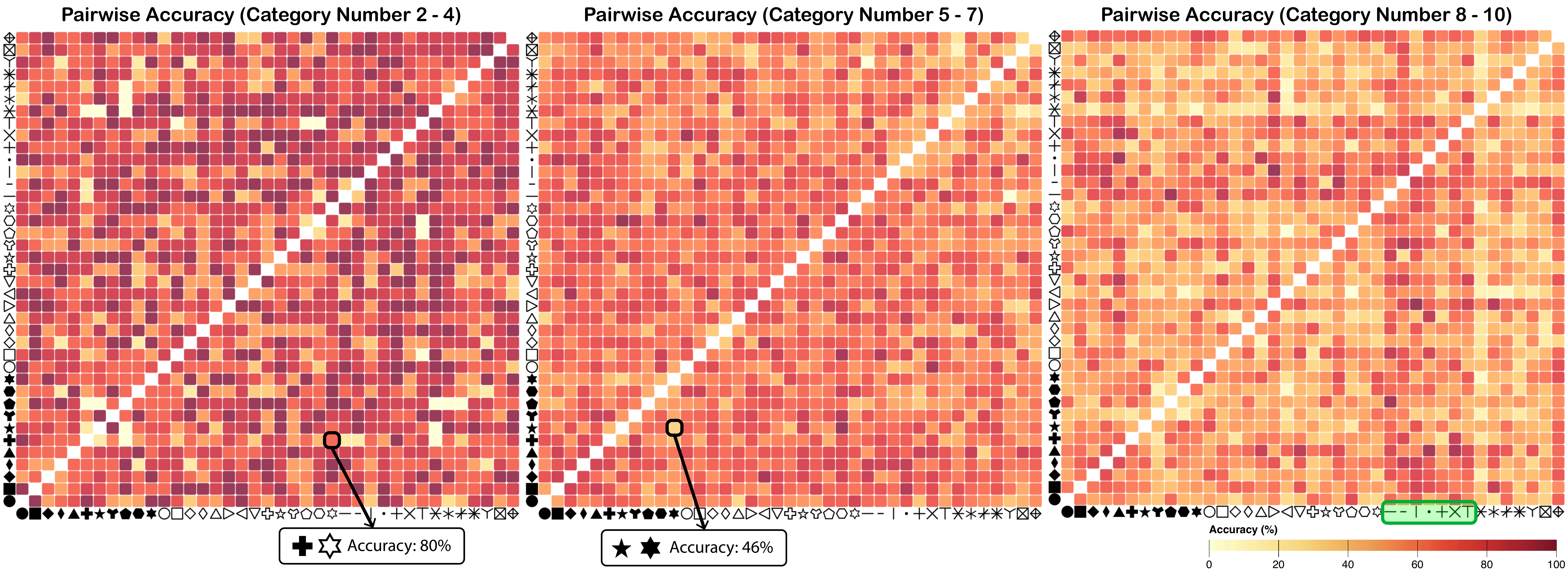} 
\vspace{-1.5em}
\caption{The pairwise accuracy matrices separated by low (2-4), middle (5-7), and high (8-10) category numbers. The color maps range from 0-100\% accuracy rate for correlation judgment tasks in Experiment 4. There are 741 entries generated from pairwise combinations from 39 shapes. Each entry represents the accuracy rate for a certain pair, like the filled plus shape
\begingroup\normalfont
  \includegraphics[height=7pt]{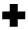}
\endgroup
and the unfilled six-pointed star
\begingroup\normalfont
  \includegraphics[height=7pt]{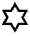}
\endgroup
has 80\% accuracy in the small number of categories matrix, and the filled five-point star
\begingroup\normalfont
  \includegraphics[height=7pt]{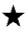}
\endgroup
and the filled six-pointed star
\begingroup\normalfont
  \includegraphics[height=7pt]{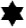}
\endgroup
has an average of 46\% accuracy in the middle number of categories matrix. For these pairwise matrices with interactive tooltip and accuracy, please refer \href{https://display-shape-results.web.app/pair.html?group=all}{here.}}
\vspace{-1em}
\label{fig:acc-pairwise}
\end{figure*}

\subsubsection{Procedure}
We applied the same general procedure as Experiments 1 and 2. Each participant completed 54 formal trials and 3 engagement checks with the target task (``identify which category is the most correlated''). Trials were displayed randomly and sequentially. One of the 15 task groups was randomly assigned to each participant at the beginning of the study, with ten valid responses collected for each group.

\subsubsection{Participants}
We recruited 168 participants via MTurk. 18 participants who did not pass the engagement checks were excluded. We analyzed the data from 150 participants (99 male, 51 female; 20--65 years of age), \add{with 10 participants in each task group.} The recruiting criteria remained the same as in the first two experiments. The study took an average of 15 minutes to complete, \add{and each participant was compensated \$2.50.}

\subsubsection{Analysis}
We used pairwise accuracy as the main dependent measure.
The pairwise accuracy was calculated based on the accuracy of every scatterplot in which \add{each} pair \add{combination} appeared. \add{For instance, if a scatterplot with circle, triangle, and square was answered correctly, the pairs circle-triangle, circle-square and triangle-square will get a correct count. We calculate the pairwise accuracy by taking each certain pair's correct count divided by how many trials the pair appeared in.}
We analyzed the resulting data using an ANOVA with category number and shape pair as independent variables.
\autoref{tab:exp4} summarizes our results. 

\begin{table}[htbp] 
\vspace{-0.5em}
\centering
\caption{ANOVA results for category number and shape pairs. Significant effects are indicated by \textbf{bold} text.}
\vspace{-0.5em}
\includegraphics[width=\linewidth]{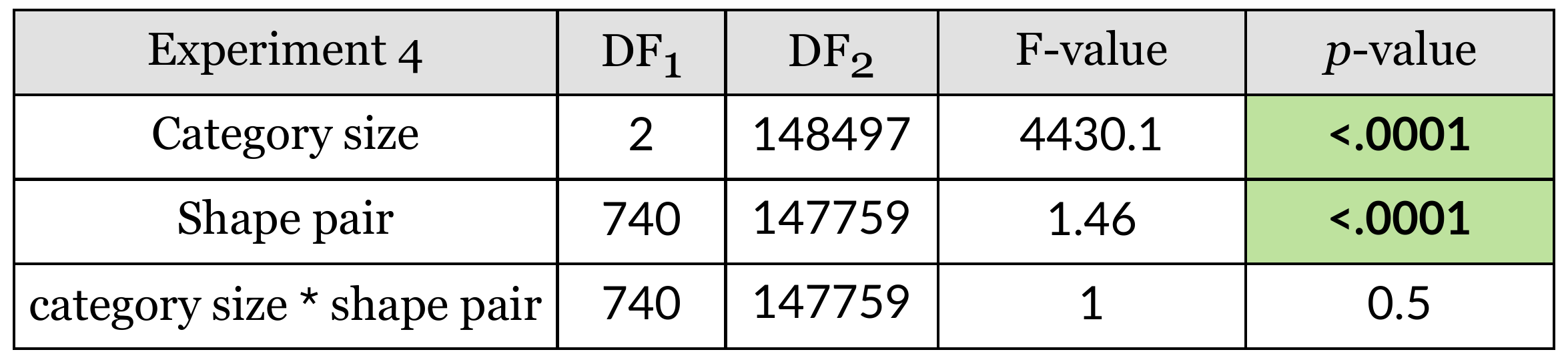} 
\label{tab:exp4}
\vspace{-2em}
\end{table}

\subsection{Results}

Our analysis revealed significant effects of both category sizes ($F(2, 148497) = 4430.1, p <.0001$) and specific shape pairs ($F(740, 147759) = 1.46, p <.0001$) on 
correlation comparison.
To better examine these results, we categorize the pairwise accuracy between shape pairs into low (2-4), middle (5-7), and high (8-10) category numbers following prior studies~\cite{tseng2023evaluating}.

\begin{figure}[t] 
\centering
\includegraphics[width=0.35\textwidth]{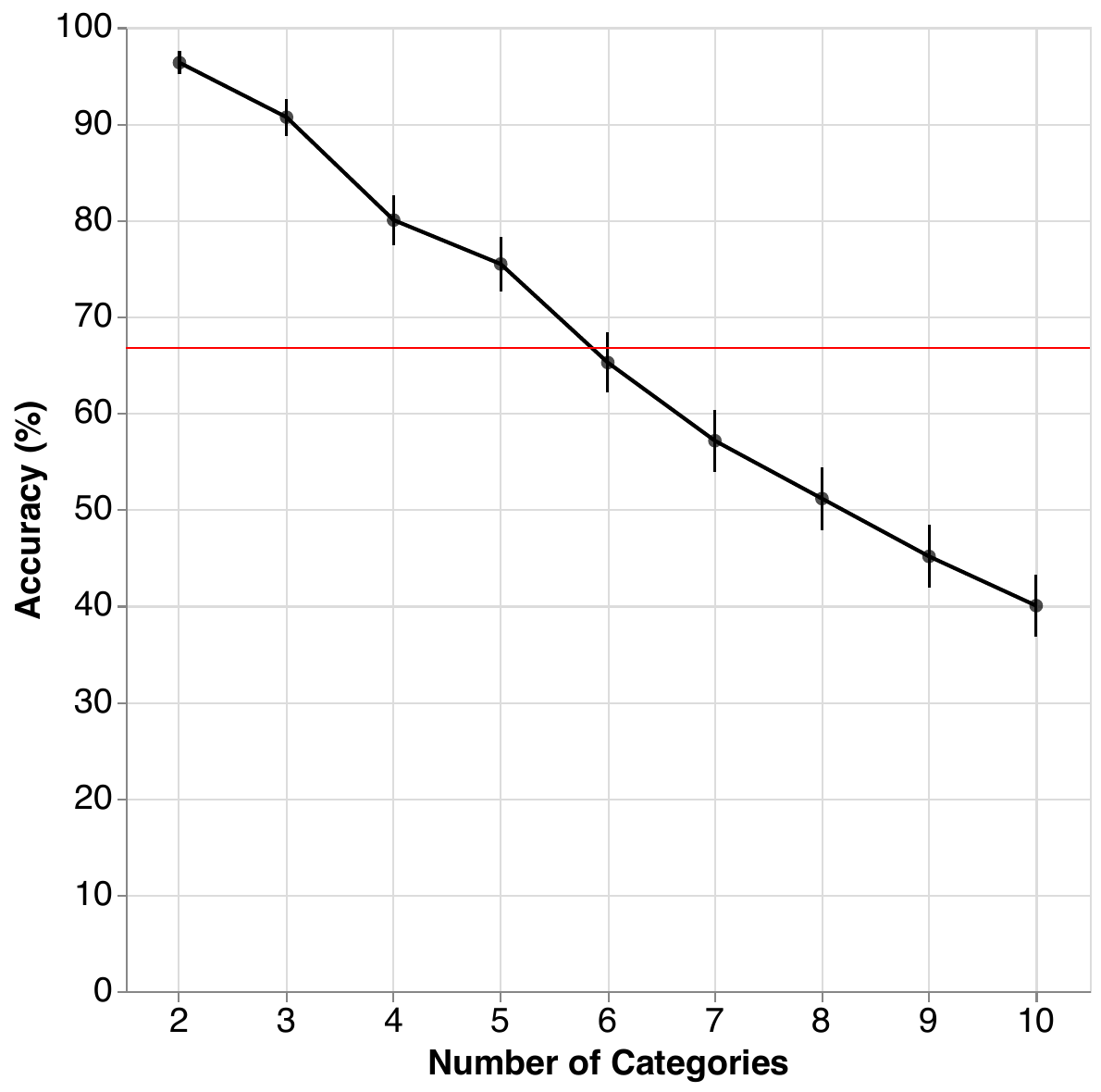} 
\vspace{-0.5em}
\caption{Overall accuracy on correlation judgment tasks separated by category numbers in Experiment 4. Accuracy consistently drops as the number of categories increases, ranging from 96\% to 40\%. The average accuracy is 66.8\%. Error bars represent 95\% confidence intervals.} 
\vspace{-1.5em}
\label{fig:acc-exp4}
\end{figure}

We summarized these results using pairwise heatmaps in \autoref{fig:acc-pairwise}.
The result reveals that the overall pairwise accuracy decreases with increasing category numbers (\autoref{fig:acc-exp4}).
However, performance and changes in performance significantly differed between different shape pairs. For example, 
open shapes with lower densities 
\begingroup\normalfont
  \includegraphics[height=7pt]{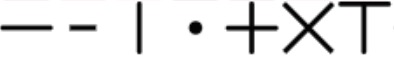}
\endgroup (in the green box) achieved relatively high pairwise accuracy at high category numbers, but accuracy did not noteably increase at lower numbers.
This robustness may help explain the performance of categorical palettes from Excel and Matlab in \autoref{fig:acc-palettes}(b) and (d), which contain several of these shapes.
The 
\begingroup\normalfont
  \includegraphics[height=7pt]{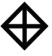}
\endgroup shape (the last one on the heatmap) achieved a relatively high pairwise accuracy overall
robust to increasing category numbers.
The accuracy of most filled and unfilled shapes largely, but not universally, depended on category numbers.

Our results support \textbf{H$_{4}$}: we found \add{\textbf{shape pairs significantly impact the accuracy of correlation comparison.}} However, aside from some benefits from low-density open shapes, we found no other notable geometric features predicting performance or robustness. 
We also found that pairwise differences alone were insufficient to fully explain performance. As the number of categories increased, performance decreased for most pairs. However, this performance decrease was non-uniform, suggesting that shapes outside of each pair interact in complex ways to change task performance. Features like shape similarity, clutter, and class overlap all likely influence palette performance.

%% file: sec-tool.tex
\subsection{Pairwise Shape Difference Model}
Since our experimental results show that shape type (Experiment 1) and geometric features (Experiment 4) are insufficient to 
predict palette performance and failed to find evidence of consistent design heuristics (Experiments 2 and 3), we made our experimental results actionable by integrating them into a model for recommending palettes based on performance. 
We generated four pairwise accuracy matrices from 14,850 samples from Experiment 4, three matrices separated by category number size (\autoref{fig:acc-pairwise}) 
and a global pairwise accuracy matrix (\autoref{fig:teaser}). \add{Given a shape set and target category number,} the model first scores a selected palette by considering the accuracy scores of the pairs in the appropriate distance matrix, as determined by the target category number and in the global matrix. After retrieving all pairwise scores \add{for every shape combination palette}, the top ten palettes are further differentiated by their overall mean pairwise accuracy computed using the data from Experiments 1 and 2, which offer a sparser set of pairwise scores compared to Experiment 4. The resulting scores generate a ranked list of shape palettes for a given number of categories, where higher ranks are likely to support better analyses. 

\subsection{Cross-Measure Validation using Pairwise Shape Model}
Our model can 
both assist shape palette design (see \autoref{sub-sec-tool}) and predict shape palette performance. For instance, our model can compare performance between several shape palettes, outputting a ranked set of palettes based on predicted performance. 
We validate our model's ability to rank palettes using the ground truth accuracy data from the tested palettes in Experiment 4 as a preliminary exploratory cross-measure validation. Specifically, we explored whether the palette rankings within a given category number consistently corresponded to ground truth performance across all category numbers.  

\begin{figure}[t] 
\centering
\includegraphics[width=0.45\textwidth]{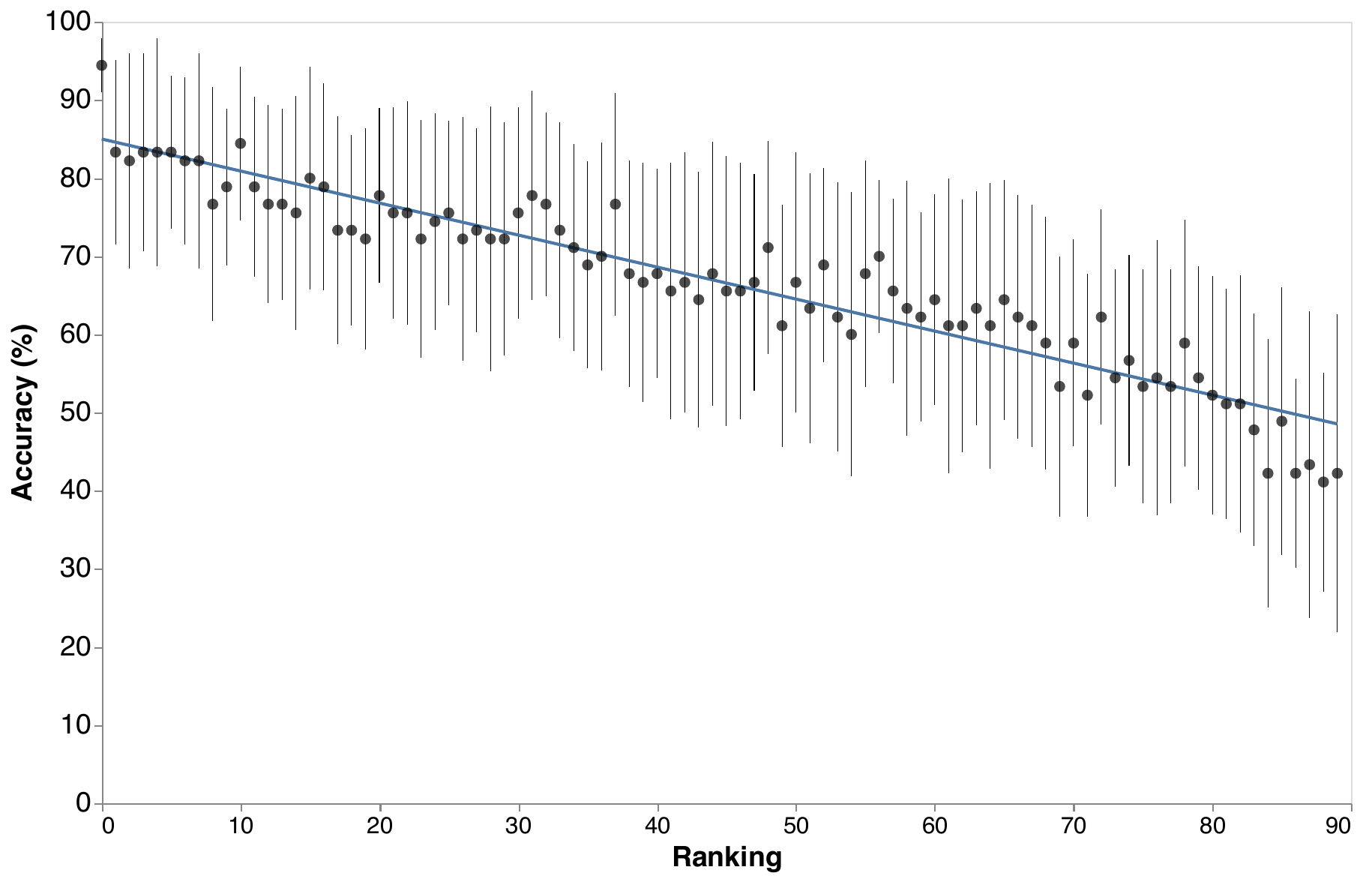} 
\vspace{0.5em}
\caption{The cross-measure validation results between our model predicted ranking and human accuracy results in Experiment 4. 
We used our model to predict the ranking within each category number, with output rankings ranging from 0-90, a higher ranking represented a better performance in our model. We calculated the average accuracy of correlation judgment tasks of categories 2-10 with those having the same ranking position. 95\% CI reported for the accuracy for categories 2-10 at that rank. The regression line shows an overall descending trend across categories with lower rankings within a category.}
\vspace{-2.0em}
\label{fig:corss}
\end{figure}

Experiment 4 tested 90 distinct shape combinations for each category number, and each combination received 10 responses. We used our model to rank the 90 combinations within each category number and calculated the average accuracy of each combination as ground truth performance.  \autoref{fig:corss} shows the relationship between the rank (1--90) within a given category number and the mean ground truth accuracy across all numbers of categories. 
The consistent downward trend shows that our rank prediction aligns well with actual accuracy: the lower the within-category performance ranking our model predicted, the worse the corresponding shape combinations performed across all tested category numbers.
We found a strong correlation between the predicted ranks and mean performance ($r=0.96, p <.0001$). 
These results provide preliminary evidence that our model can faithfully 
predict performance across palettes; however, it focuses on comparing ranked predictions within our dataset. Formal model validation across a broader range of data is important for future work. 

\subsection{Design Tool for Shape Palettes} \label{sub-sec-tool}

We implemented our model in a web-based shape recommendation tool (see \textcolor{blue}{\href{https://shape-it-up-eec5e.web.app/}{web resource}}\footnote{\url{https://shape-it-up-eec5e.web.app/}}).
Visualization tools commonly provide a set of shapes for users to encode categorical data in a scatterplot. However, most tools require users to either use a predefined palette or input their shape selection independently.
Both approaches lack grounding in empirical studies. Further, our results in Experiment 3 suggest that users exhibit diverse shape preferences that using predefined palettes cannot account for.

Our shape recommendation 
tool allows users to choose seed shapes to start a palette and then suggests an optimally scoring palette for a specified number of categories that include the selected shapes. As shown in \autoref{fig:teaser}, users can select any number of desired seed shapes from a predefined shape pool and then select the number of shapes in the final palette.  The tool recommends a high-performing palette based on these parameters. If the number of preselected shapes is larger than the target number of categories, the system will design a palette from the preselected shapes that optimizes predicted performance based on our model. Otherwise, the tool will use all of the preselected shapes and add new candidate shapes based on our scoring model. Two previews of scatterplots using generated shape sets give users a sense of what the resulting design will look like. 
The tool provides three main features:

\vspace{6pt}
\noindent\textit{1.} \textit{
Diverse shape types.} Users can select from any of the 39 shapes found in our survey of past research and commercial tools. The set of possible shapes is shown in a selection window.  

\noindent\textit{2.} \textit{Allow users to select preferred shapes and swap the output shapes.} 
Users can select their 
preferred shapes by clicking them in the shape selection window. Once a shape palette is generated, users can 
remove the shapes they do not want from the generated palette by selecting them in the output window, 
and the tool will replace them with new, high-scoring shapes. This feature provides flexibility for using the tool to satisfy preferences and ensure high performance concurrently.

\noindent\textit{3.} \textit{Providing a set of shapes to match the required category number.} 
The underlying selection model is based on 380 participants' responses with from 2 to 10 categories collected over two supporting tasks. By integrating Experiments 1, 2, and 4, we calculate the pair scores based on six pairwise accuracy matrices from two tasks and sensitive to a user's target data needs. 

To generate sets of shapes, we compare palette scores based on our pairwise shape difference model. Since it is computationally prohibitive to calculate scores for every combination from 39 shapes given the combinatorial number of considerations, we fixed the calculation time to return the highest scoring palette found within $\sim$5 secs. This allows us to compare around 400 random shape combinations for ten categories and is similar to the construction process in Colorgorical~\cite{gramazio2016colorgorical}.

%% file: sec-discussion.tex
Our findings 
offer empirical insight into categorical shape encoding perception for multiclass scatterplots. 
These findings contribute to a growing literature on shape perception in visualization, principles of effective categorical encoding, and methods for generating visualization designs grounded in empirical data. 

Despite being a common default method for encoding categorical data, shape palettes 
are far less studied than color~\cite{burlinson2017open, smart2019measuring}.
Understanding shape is important in part because 
it offers an intuitive means to represent discrete data points, 
especially when color cannot be used either due to technical constraints or to accommodate color vision deficiencies~\cite{ware2009quantitative, elavsky2022accessible,burlinson2017open}. Systems like diatoms \cite{brehmer2021generative} explore how generative and designerly approaches might create novel shape palettes; however, while shape features have long been a topic of study within scientific visualization for comparing complex shapes \cite{heider2011local}, we lack insight into the features that support effective shape palette design. Our results indicate that the features that characterize effective shape palettes are complex. We found few discernible patterns in what shapes align to create effective palettes and high performance variance between professional palettes. While our model aims to make our results actionable, the lack of discernible patterns suggests the need for more extensive studies to understand the characteristics of effective shape palette design. 

Our results offer additional insight into prior results.
For example, we confirm Burlinson et al.'s finding that open shapes provide worse averaging performance, but, in contrast to their results, integrating multiple shape types can increase performance \cite{burlinson2017open}. This difference is likely due to testing a larger number of categories, as within-category shapes tend to become more geometrically similar as more shapes are sampled. Adding type variation can likely increase distinguishability as the number of categories scales.  
We also confirm 
increasing category numbers decreases performance, 
further confirming existing heuristics for 
palette design that suggests designers must balance the number of encoded categories and ease of data perception~\cite{tseng2023evaluating,tseng2024revisiting,haroz2012capacity}.

Unlike prior experiments, however, we found that people struggled significantly 
estimating the average value in a set of data points when using shape palettes for a large number of categories.
We used a comparable stimulus set to 
Tseng et al. \cite{tseng2023evaluating}, who found that when using color, people still achieve more than 70\% accuracy at 10 categories. We found some shape palettes and types exhibited less than 50\% accuracy at 10 categories. 
Similarly, our results indicate a lower performance in correlation estimation studies of scatterplots~\cite{rensink2010perception, harrison2014ranking, kay2015beyond}, suggesting that visual cues across multiple categories can significantly impact the accuracy of correlation estimates.
These results indicate a further need to better understand shape palette design in similar ways to color palettes to help predict when performance will falter and to create palettes that are more robust to larger numbers of categories. 

\subsection{Design Implications}

While our results do not necessarily illuminate comprehensive heuristics, we did find evidence of promising strategies for shape palette design. Based on the results, we recommend: 
\begin{itemize}
    \item Filled+open shapes effectively supports 
    fewer than six categories. 
    \item 
    For six or more categories, 
    a mix of different shape types or a bias towards low-density open shapes may provide more robust performance. 
    \item Selecting shapes from different types tends to improve performance compared to using any single type.
\end{itemize}

Our results indicate that default palettes may also be problematic: 
we found significant variability in the performance of design-crafted shape palettes with 
largest accuracy differences exceeding 20\% for some category numbers.
Experiments 2 and 3 show 
that even experts 
lack a uniform approach to selecting palette shapes,
pointing to significant variance in individual preferences.
This discrepancy highlights the need for empirical methods to guide shape selection.

To make our results more actionable, we introduce a tool
that can aid in the  design of effective shape palettes.
Our approach aims to allow designers to select shapes based on their preferences and specific needs while providing task-oriented recommendations within those constraints. Given the diversity of possible shapes and tasks, we hope to expand this model as new research continues to investigate the role of shape in categorical data encoding and graphical perception.

\subsection{Limitations and Future Work}

We employed a series of experiments to investigate the characteristics of effective shape palettes. Our studies focused on shape types, shape palettes, and pairwise distances as primary variables. However, future work should investigate a range of additional factors. 
For example, shape encodings often use differing orientations of a certain shape to represent different categories (e.g., the triangles in the Tableau palette). Moreover, 
designers may benefit from understanding the accuracy of shapes' orientations on a larger range of tasks, geometries, and scatterplot types. For example, orientation may operate differently in an abstract scatterplot compared to one overlaid on a map, where orientation may imply perceived direction. Future work should explore how orientation impacts 
performance across different tasks. 

Our experiments employed mean and correlation judgment tasks to model the perceptual distance between shape pairs. Future work should investigate how different tasks used in multiclass scatterplots interplay with shape performance.
For example, certain tasks, such as extrema detection, may require more direct shape comparison. \add{Both our study and prior work \cite{tseng2023evaluating, gleicher2013perception} focused on comparing y values, yet future work should also consider how palettes affect tasks involving the x-dimension.}
Integrating data from these tasks into our target model would further strengthen it to account for a broader range of tasks and could contribute to a more universal shape space model for palette design.
Preference-based tasks, such as Experiment 3, also may further strengthen the model by integrating more aesthetic elements into palette design. Future work should consider how to best weight these factors. 

We generated stimuli using a fixed mark size and stroke width for every shape. Varying sizes can influence mark discriminability \cite{smart2019measuring}. For example, we found that different mark sizes, like the dot in Matlab or the shorter line in Excel, have different performance. 
However, we do not anticipate the performance differences using varied mark sizes given the limited effect of size on shape perception in prior visualization studies \cite{smart2019measuring}. 
Exploring the interaction between shape and other visual channels, such as size and color, specifically for palette design is important for future work to understand the robustness of our model. 

We limited our shape pool to the most common shapes from widely used visualization tools. While this still led to a significantly larger set of shapes than prior work (39 total), it may not capture all shape features sufficiently. Unlike colormaps, where designers can use computational models of hue or lightness to construct a color palette, shapes have a more diverse range of visual features. While we aimed to capture a foundational understanding of shape palettes based on existing design approaches, future work should more extensively explore variation across a wider range of shape features and how feature models designed for more complex shapes may help model palette design with simple shapes, such as those used in multiclass scatterplots.

Our study did not account for some other psychological factors of shape, such as aesthetics
\cite{schloss2011aesthetic}, artistic 
theories~\cite{szafir2023visualization, bujack2017good}, and attentional weight~\cite{hong2021weighted}. 
Other factors like the cognitive load and capacity limit of categories may also affect users’ ability and capacity to process categorical visualizations~\cite{haroz2012capacity, tseng2023evaluating}.
Future studies should explore how various psychological factors involving shape affect palette design.

%% file: sec-conclusion.tex
We investigated key factors involved in constructing effective shape palettes for multiclass scatterplots. Our findings indicate that shape is a complex attribute that cannot be adequately assessed by broad types and that even professional designs exhibited significant performance variation. 
We discovered that shape pairs significantly influence people's ability to differentiate between classes. Consequently, we developed a pairwise distance matrix to model the perceptual differences between shapes. Moreover, we implemented our findings in a palette design tool, empowering designers to make informed decisions when selecting shapes. This tool not only offers shape palettes backed by empirical research but also allows for flexibility in integrating user preferences. Our research sheds new light on shape encodings for categorical data and the lack of actionable design guidance for creating such palettes. We hope our work will inspire future studies aimed at establishing broader principles for effective categorical visualization.

%% file: main.bbl
\begin{thebibliography}{10}

\bibitem{abbas2019clustme}
M.~Abbas, M.~Aupetit, M.~Sedlmair, and H.~Bensmail.
\newblock Clustme: A visual quality measure for ranking monochrome scatterplots based on cluster patterns.
\newblock {\em Comput. Graph. Forum}, 38(3):225--236, 2019. \href{https://doi.org/10.1111/cgf.13684}
{doi: {{%
10\hspace{.1pt}\discretionary{.}{%
}{.}\hspace{.4pt}1111\discretionary{/}{%
}{/}cgf\hspace{.1pt}\discretionary{.}{%
}{.}\hspace{.4pt}13684}}}


\bibitem{ali2013effect}
N.~Ali and D.~Peebles.
\newblock The effect of gestalt laws of perceptual organization on the comprehension of three-variable bar and line graphs.
\newblock {\em Human factors}, 55(1):183--203, 2013. \href{https://doi.org/10.1177/0018720812452592}
{doi: {{%
10\hspace{.1pt}\discretionary{.}{%
}{.}\hspace{.4pt}1177\discretionary{/}{%
}{/}0018720812452592}}}


\bibitem{baker2018abstract}
N.~Baker and P.~J. Kellman.
\newblock Abstract shape representation in human visual perception.
\newblock {\em Journal of Experimental Psychology: General}, 147(9):1295, 2018. \href{https://doi.org/10.1037/xge0000409}
{doi: {{%
10\hspace{.1pt}\discretionary{.}{%
}{.}\hspace{.4pt}1037\discretionary{/}{%
}{/}xge0000409}}}


\bibitem{6064996}
M.~Bostock, V.~Ogievetsky, and J.~Heer.
\newblock D³ data-driven documents.
\newblock {\em IEEE Trans. Vis. Comput. Graph.}, 17(12):2301--2309, 2011.

\bibitem{brath2010multiple}
R.~Brath.
\newblock Multiple shape attributes in information visualization: Guidance from prior art and experiments.
\newblock In {\em 2010 14th International Conference Information Visualisation}, pp. 433--438. IEEE, 2010. \href{https://doi.org/10.1109/IV.2010.66}
{doi: {{%
10\hspace{.1pt}\discretionary{.}{%
}{.}\hspace{.4pt}1109\discretionary{/}{%
}{/}IV\hspace{.1pt}\discretionary{.}{%
}{.}\hspace{.4pt}2010\hspace{.1pt}\discretionary{.}{%
}{.}\hspace{.4pt}66}}}


\bibitem{brehmer2021generative}
M.~Brehmer, R.~Kosara, and C.~Hull.
\newblock Generative design inspiration for glyphs with diatoms.
\newblock {\em IEEE Transactions on Visualization and Computer Graphics}, 28(1):389--399, 2021. \href{https://doi.org/10.1109/tvcg.2021.3114792}
{doi: {{%
10\hspace{.1pt}\discretionary{.}{%
}{.}\hspace{.4pt}1109\discretionary{/}{%
}{/}tvcg\hspace{.1pt}\discretionary{.}{%
}{.}\hspace{.4pt}2021\hspace{.1pt}\discretionary{.}{%
}{.}\hspace{.4pt}3114792}}}


\bibitem{bujack2017good}
R.~Bujack, T.~L. Turton, F.~Samsel, C.~Ware, D.~H. Rogers, and J.~Ahrens.
\newblock The good, the bad, and the ugly: A theoretical framework for the assessment of continuous colormaps.
\newblock {\em IEEE Trans. Vis. Comput. Graph.}, 24(1):923--933, 2017. \href{https://doi.org/10.1109/TVCG.2017.2743978}
{doi: {{%
10\hspace{.1pt}\discretionary{.}{%
}{.}\hspace{.4pt}1109\discretionary{/}{%
}{/}TVCG\hspace{.1pt}\discretionary{.}{%
}{.}\hspace{.4pt}2017\hspace{.1pt}\discretionary{.}{%
}{.}\hspace{.4pt}2743978}}}


\bibitem{burlinson2017open}
D.~Burlinson, K.~Subramanian, and P.~Goolkasian.
\newblock Open vs. closed shapes: New perceptual categories?
\newblock {\em IEEE Trans. Vis. Comput. Graph.}, 24(1):574--583, 2017. \href{https://doi.org/10.1109/TVCG.2017.2745086}
{doi: {{%
10\hspace{.1pt}\discretionary{.}{%
}{.}\hspace{.4pt}1109\discretionary{/}{%
}{/}TVCG\hspace{.1pt}\discretionary{.}{%
}{.}\hspace{.4pt}2017\hspace{.1pt}\discretionary{.}{%
}{.}\hspace{.4pt}2745086}}}


\bibitem{casperson1950visual}
R.~C. Casperson.
\newblock The visual discrimination of geometric forms.
\newblock {\em Journal of Experimental Psychology}, 40(5):668, 1950. \href{https://doi.org/10.1037/h0055220}
{doi: {{%
10\hspace{.1pt}\discretionary{.}{%
}{.}\hspace{.4pt}1037\discretionary{/}{%
}{/}h0055220}}}


\bibitem{dai2022visual}
L.~Dai, K.~Zhang, X.~S. Zheng, R.~R. Martin, Y.~Li, and J.~Yu.
\newblock Visual complexity of shapes: a hierarchical perceptual learning model.
\newblock {\em The Visual Computer}, pp. 1--14, 2022. \href{https://doi.org/10.1007/s00371-020-02023-z}
{doi: {{%
10\hspace{.1pt}\discretionary{.}{%
}{.}\hspace{.4pt}1007\discretionary{/}{%
}{/}s00371\discretionary{%
}{-}{-}020\discretionary{%
}{-}{-}02023\discretionary{%
}{-}{-}z}}}


\bibitem{demiralp2014learning}
{\c{C}}.~Demiralp, M.~S. Bernstein, and J.~Heer.
\newblock Learning perceptual kernels for visualization design.
\newblock {\em IEEE Trans. Vis. Comput. Graph.}, 20(12):1933--1942, 2014. \href{https://doi.org/10.1109/TVCG.2014.2346978}
{doi: {{%
10\hspace{.1pt}\discretionary{.}{%
}{.}\hspace{.4pt}1109\discretionary{/}{%
}{/}TVCG\hspace{.1pt}\discretionary{.}{%
}{.}\hspace{.4pt}2014\hspace{.1pt}\discretionary{.}{%
}{.}\hspace{.4pt}2346978}}}


\bibitem{deutsch1955theory}
J.~Deutsch.
\newblock A theory of shape recognition.
\newblock {\em British Journal of Psychology}, 46(1):30, 1955.

\bibitem{ebert2000procedural}
D.~S. Ebert, R.~M. Rohrer, C.~D. Shaw, P.~Panda, J.~M. Kukla, and D.~A. Roberts.
\newblock Procedural shape generation for multi-dimensional data visualization.
\newblock {\em Computers \& Graphics}, 24(3):375--384, 2000. \href{https://doi.org/10.1007/978-3-7091-6803-5_1}
{doi: {{%
10\hspace{.1pt}\discretionary{.}{%
}{.}\hspace{.4pt}1007\discretionary{/}{%
}{/}978\discretionary{%
}{-}{-}3\discretionary{%
}{-}{-}7091\discretionary{%
}{-}{-}6803\discretionary{%
}{-}{-}5\_1}}}


\bibitem{elavsky2022accessible}
F.~Elavsky, C.~Bennett, and D.~Moritz.
\newblock How accessible is my visualization? evaluating visualization accessibility with chartability.
\newblock {\em Computer Graphics Forum}, 41(3):57--70, 2022. \href{https://doi.org/10.1111/cgf.14522}
{doi: {{%
10\hspace{.1pt}\discretionary{.}{%
}{.}\hspace{.4pt}1111\discretionary{/}{%
}{/}cgf\hspace{.1pt}\discretionary{.}{%
}{.}\hspace{.4pt}14522}}}


\bibitem{elliott2020design}
M.~A. Elliott, C.~Nothelfer, C.~Xiong, and D.~A. Szafir.
\newblock A design space of vision science methods for visualization research.
\newblock {\em IEEE Trans. Vis. Comput. Graph.}, 2020. \href{https://doi.org/10.1109/TVCG.2020.3029413}
{doi: {{%
10\hspace{.1pt}\discretionary{.}{%
}{.}\hspace{.4pt}1109\discretionary{/}{%
}{/}TVCG\hspace{.1pt}\discretionary{.}{%
}{.}\hspace{.4pt}2020\hspace{.1pt}\discretionary{.}{%
}{.}\hspace{.4pt}3029413}}}


\bibitem{gleicher2013perception}
M.~Gleicher, M.~Correll, C.~Nothelfer, and S.~Franconeri.
\newblock Perception of average value in multiclass scatterplots.
\newblock {\em IEEE Trans. Vis. Comput. Graph.}, 19, 2013. \href{https://doi.org/10.1109/TVCG.2013.183}
{doi: {{%
10\hspace{.1pt}\discretionary{.}{%
}{.}\hspace{.4pt}1109\discretionary{/}{%
}{/}TVCG\hspace{.1pt}\discretionary{.}{%
}{.}\hspace{.4pt}2013\hspace{.1pt}\discretionary{.}{%
}{.}\hspace{.4pt}183}}}


\bibitem{goldstone2010categorical}
R.~L. Goldstone and A.~T. Hendrickson.
\newblock Categorical perception.
\newblock {\em Wiley Interdiscip. Rev.: Cogn. Sci.}, 1(1):69--78, 2010. \href{https://doi.org/10.1002/wcs.26}
{doi: {{%
10\hspace{.1pt}\discretionary{.}{%
}{.}\hspace{.4pt}1002\discretionary{/}{%
}{/}wcs\hspace{.1pt}\discretionary{.}{%
}{.}\hspace{.4pt}26}}}


\bibitem{gramazio2016colorgorical}
C.~C. Gramazio, D.~H. Laidlaw, and K.~B. Schloss.
\newblock Colorgorical: Creating discriminable and preferable color palettes for information visualization.
\newblock {\em IEEE Trans. Vis. Comput. Graph.}, 23(1):521--530, 2016. \href{https://doi.org/10.1109/TVCG.2016.2598918}
{doi: {{%
10\hspace{.1pt}\discretionary{.}{%
}{.}\hspace{.4pt}1109\discretionary{/}{%
}{/}TVCG\hspace{.1pt}\discretionary{.}{%
}{.}\hspace{.4pt}2016\hspace{.1pt}\discretionary{.}{%
}{.}\hspace{.4pt}2598918}}}


\bibitem{haroz2012capacity}
S.~Haroz and D.~Whitney.
\newblock How capacity limits of attention influence information visualization effectiveness.
\newblock {\em IEEE Trans. Vis. Comput. Graph.}, 18(12), 2012. \href{https://doi.org/10.1109/TVCG.2012.233}
{doi: {{%
10\hspace{.1pt}\discretionary{.}{%
}{.}\hspace{.4pt}1109\discretionary{/}{%
}{/}TVCG\hspace{.1pt}\discretionary{.}{%
}{.}\hspace{.4pt}2012\hspace{.1pt}\discretionary{.}{%
}{.}\hspace{.4pt}233}}}


\bibitem{2020NumPy-Array}
C.~R. Harris, K.~J. Millman, S.~J. van~der Walt, R.~Gommers, P.~Virtanen, D.~Cournapeau, E.~Wieser, J.~Taylor, S.~Berg, N.~J. Smith, R.~Kern, M.~Picus, S.~Hoyer, M.~H. van Kerkwijk, M.~Brett, A.~Haldane, J.~Fernández~del Río, M.~Wiebe, P.~Peterson, P.~Gérard-Marchant, K.~Sheppard, T.~Reddy, W.~Weckesser, H.~Abbasi, C.~Gohlke, and T.~E. Oliphant.
\newblock Array programming with {NumPy}.
\newblock {\em Nature}, 585:357–362, 2020. \href{https://doi.org/10.1038/s41586-020-2649-2}
{doi: {{%
10\hspace{.1pt}\discretionary{.}{%
}{.}\hspace{.4pt}1038\discretionary{/}{%
}{/}s41586\discretionary{%
}{-}{-}020\discretionary{%
}{-}{-}2649\discretionary{%
}{-}{-}2}}}


\bibitem{harrison2014ranking}
L.~Harrison, F.~Yang, S.~Franconeri, and R.~Chang.
\newblock Ranking visualizations of correlation using weber's law.
\newblock {\em IEEE transactions on visualization and computer graphics}, 20(12):1943--1952, 2014. \href{https://doi.org/10.1109/TVCG.2014.2346979}
{doi: {{%
10\hspace{.1pt}\discretionary{.}{%
}{.}\hspace{.4pt}1109\discretionary{/}{%
}{/}TVCG\hspace{.1pt}\discretionary{.}{%
}{.}\hspace{.4pt}2014\hspace{.1pt}\discretionary{.}{%
}{.}\hspace{.4pt}2346979}}}


\bibitem{harrower2003colorbrewer}
M.~Harrower and C.~A. Brewer.
\newblock Colorbrewer. org: an online tool for selecting colour schemes for maps.
\newblock {\em Cartogr. J.}, 40(1):27--37, 2003. \href{https://doi.org/10.1179/000870403235002042}
{doi: {{%
10\hspace{.1pt}\discretionary{.}{%
}{.}\hspace{.4pt}1179\discretionary{/}{%
}{/}000870403235002042}}}


\bibitem{heer2012color}
J.~Heer and M.~Stone.
\newblock Color naming models for color selection, image editing and palette design.
\newblock In {\em Proc. ACM Hum. Factors Comput. Syst. (CHI)}, pp. 1007--1016, 2012. \href{https://doi.org/10.1145/2207676.2208547}
{doi: {{%
10\hspace{.1pt}\discretionary{.}{%
}{.}\hspace{.4pt}1145\discretionary{/}{%
}{/}2207676\hspace{.1pt}\discretionary{.}{%
}{.}\hspace{.4pt}2208547}}}


\bibitem{heider2011local}
P.~Heider, A.~Pierre-Pierre, R.~Li, and C.~Grimm.
\newblock Local shape descriptors, a survey and evaluation.
\newblock In {\em Proceedings of the 4th Eurographics Conference on 3D Object Retrieval},  8 pages. The Eurographics Association, 2011. \href{https://doi.org//10.2312/3DOR/3DOR11/049-056}
{doi: {{%
\discretionary{/}{%
}{/}10\hspace{.1pt}\discretionary{.}{%
}{.}\hspace{.4pt}2312\discretionary{/}{%
}{/}3DOR\discretionary{/}{%
}{/}3DOR11\discretionary{/}{%
}{/}049\discretionary{%
}{-}{-}056}}}


\bibitem{hong2021weighted}
M.-H. Hong, J.~K. Witt, and D.~A. Szafir.
\newblock The weighted average illusion: Biases in perceived mean position in scatterplots.
\newblock {\em IEEE Trans. Vis. Comput. Graph.}, 28(1):987--997, 2021. \href{https://doi.org/10.1109/tvcg.2021.3114783}
{doi: {{%
10\hspace{.1pt}\discretionary{.}{%
}{.}\hspace{.4pt}1109\discretionary{/}{%
}{/}tvcg\hspace{.1pt}\discretionary{.}{%
}{.}\hspace{.4pt}2021\hspace{.1pt}\discretionary{.}{%
}{.}\hspace{.4pt}3114783}}}


\bibitem{Hunter:2007}
J.~D. Hunter.
\newblock Matplotlib: A 2d graphics environment.
\newblock {\em Computing in Science \& Engineering}, 9(3):90--95, 2007. \href{https://doi.org/10.1109/MCSE.2007.55}
{doi: {{%
10\hspace{.1pt}\discretionary{.}{%
}{.}\hspace{.4pt}1109\discretionary{/}{%
}{/}MCSE\hspace{.1pt}\discretionary{.}{%
}{.}\hspace{.4pt}2007\hspace{.1pt}\discretionary{.}{%
}{.}\hspace{.4pt}55}}}


\bibitem{plotly}
P.~T. Inc.
\newblock Collaborative data science, 2015.

\bibitem{MATLAB}
T.~M. Inc.
\newblock Matlab version: 9.13.0 (r2022b), 2022.

\bibitem{jeon2023clams}
H.~Jeon, G.~J. Quadri, H.~Lee, P.~Rosen, D.~A. Szafir, and J.~Seo.
\newblock Clams: A cluster ambiguity measure for estimating perceptual variability in visual clustering.
\newblock {\em IEEE Trans. Vis. Comput. Graph.}, 2023. \href{https://doi.org/10.1109/tvcg.2023.3327201}
{doi: {{%
10\hspace{.1pt}\discretionary{.}{%
}{.}\hspace{.4pt}1109\discretionary{/}{%
}{/}tvcg\hspace{.1pt}\discretionary{.}{%
}{.}\hspace{.4pt}2023\hspace{.1pt}\discretionary{.}{%
}{.}\hspace{.4pt}3327201}}}


\bibitem{kay2015beyond}
M.~Kay and J.~Heer.
\newblock Beyond weber's law: A second look at ranking visualizations of correlation.
\newblock {\em IEEE Trans. Vis. Comput. Graph.}, 22(1):469--478, 2015. \href{https://doi.org/10.1109/TVCG.2015.2467671}
{doi: {{%
10\hspace{.1pt}\discretionary{.}{%
}{.}\hspace{.4pt}1109\discretionary{/}{%
}{/}TVCG\hspace{.1pt}\discretionary{.}{%
}{.}\hspace{.4pt}2015\hspace{.1pt}\discretionary{.}{%
}{.}\hspace{.4pt}2467671}}}


\bibitem{kim2018assessing}
Y.~Kim and J.~Heer.
\newblock Assessing effects of task and data distribution on the effectiveness of visual encodings.
\newblock {\em Comput. Graph. Forum}, 37(3):157--167, 2018. \href{https://doi.org/10.1111/cgf.13409}
{doi: {{%
10\hspace{.1pt}\discretionary{.}{%
}{.}\hspace{.4pt}1111\discretionary{/}{%
}{/}cgf\hspace{.1pt}\discretionary{.}{%
}{.}\hspace{.4pt}13409}}}


\bibitem{kramer2017visual}
R.~S. Kramer, C.~G. Telfer, and A.~Towler.
\newblock Visual comparison of two data sets: do people use the means and the variability?
\newblock {\em Journal of Numerical Cognition}, 3(1), 2017. \href{https://doi.org/10.5964/jnc.v3i1.100}
{doi: {{%
10\hspace{.1pt}\discretionary{.}{%
}{.}\hspace{.4pt}5964\discretionary{/}{%
}{/}jnc\hspace{.1pt}\discretionary{.}{%
}{.}\hspace{.4pt}v3i1\hspace{.1pt}\discretionary{.}{%
}{.}\hspace{.4pt}100}}}


\bibitem{krzywinski2013plotting}
M.~Krzywinski and B.~Wong.
\newblock Plotting symbols: choose distinct symbols that overlap without ambiguity and communicate relationships in data.
\newblock {\em Nature methods}, 10(6):451--452, 2013. \href{https://doi.org/10.1038/nmeth.2490}
{doi: {{%
10\hspace{.1pt}\discretionary{.}{%
}{.}\hspace{.4pt}1038\discretionary{/}{%
}{/}nmeth\hspace{.1pt}\discretionary{.}{%
}{.}\hspace{.4pt}2490}}}


\bibitem{li2020validated}
A.~Y. Li, J.~C. Liang, A.~C. Lee, and M.~D. Barense.
\newblock The validated circular shape space: Quantifying the visual similarity of shape.
\newblock {\em Journal of Experimental Psychology: General}, 149(5):949, 2020. \href{https://doi.org/10.1037/xge0000693}
{doi: {{%
10\hspace{.1pt}\discretionary{.}{%
}{.}\hspace{.4pt}1037\discretionary{/}{%
}{/}xge0000693}}}


\bibitem{lin2013selecting}
S.~Lin, J.~Fortuna, C.~Kulkarni, M.~Stone, and J.~Heer.
\newblock Selecting semantically-resonant colors for data visualization.
\newblock {\em Comput. Graph. Forum}, 32(3pt4):401--410, 2013. \href{https://doi.org/10.1111/cgf.12127}
{doi: {{%
10\hspace{.1pt}\discretionary{.}{%
}{.}\hspace{.4pt}1111\discretionary{/}{%
}{/}cgf\hspace{.1pt}\discretionary{.}{%
}{.}\hspace{.4pt}12127}}}


\bibitem{loffler2008perception}
G.~Loffler.
\newblock Perception of contours and shapes: Low and intermediate stage mechanisms.
\newblock {\em Vision research}, 48(20):2106--2127, 2008. \href{https://doi.org/10.1016/j.visres.2008.03.006}
{doi: {{%
10\hspace{.1pt}\discretionary{.}{%
}{.}\hspace{.4pt}1016\discretionary{/}{%
}{/}j\hspace{.1pt}\discretionary{.}{%
}{.}\hspace{.4pt}visres\hspace{.1pt}\discretionary{.}{%
}{.}\hspace{.4pt}2008\hspace{.1pt}\discretionary{.}{%
}{.}\hspace{.4pt}03\hspace{.1pt}\discretionary{.}{%
}{.}\hspace{.4pt}006}}}


\bibitem{micallef2017towards}
L.~Micallef, G.~Palmas, A.~Oulasvirta, and T.~Weinkauf.
\newblock Towards perceptual optimization of the visual design of scatterplots.
\newblock {\em IEEE Trans. Vis. Comput. Graph.}, 23(6):1588--1599, 2017. \href{https://doi.org/10.1109/TVCG.2017.2674978}
{doi: {{%
10\hspace{.1pt}\discretionary{.}{%
}{.}\hspace{.4pt}1109\discretionary{/}{%
}{/}TVCG\hspace{.1pt}\discretionary{.}{%
}{.}\hspace{.4pt}2017\hspace{.1pt}\discretionary{.}{%
}{.}\hspace{.4pt}2674978}}}


\bibitem{msexcel}
{Microsoft Corporation}.
\newblock Microsoft excel.

\bibitem{mukherjee2021context}
K.~Mukherjee, B.~Yin, B.~E. Sherman, L.~Lessard, and K.~B. Schloss.
\newblock Context matters: A theory of semantic discriminability for perceptual encoding systems.
\newblock {\em IEEE Trans. Vis. Comput. Graph.}, 28(1):697--706, 2021. \href{https://doi.org/10.1109/tvcg.2021.3114780}
{doi: {{%
10\hspace{.1pt}\discretionary{.}{%
}{.}\hspace{.4pt}1109\discretionary{/}{%
}{/}tvcg\hspace{.1pt}\discretionary{.}{%
}{.}\hspace{.4pt}2021\hspace{.1pt}\discretionary{.}{%
}{.}\hspace{.4pt}3114780}}}


\bibitem{munzner2014visualization}
T.~Munzner.
\newblock {\em Visualization analysis and design}.
\newblock CRC press, 2014.

\bibitem{pasupathy2018visual}
A.~Pasupathy, Y.~El-Shamayleh, and D.~V. Popovkina.
\newblock Visual shape and object perception.
\newblock In {\em Oxford research encyclopedia of neuroscience}. 2018. \href{https://doi.org/10.1093/acrefore/9780190264086.013.75}
{doi: {{%
10\hspace{.1pt}\discretionary{.}{%
}{.}\hspace{.4pt}1093\discretionary{/}{%
}{/}acrefore\discretionary{/}{%
}{/}9780190264086\hspace{.1pt}\discretionary{.}{%
}{.}\hspace{.4pt}013\hspace{.1pt}\discretionary{.}{%
}{.}\hspace{.4pt}75}}}


\bibitem{quadri2022automatic}
G.~J. Quadri, J.~A. Nieves, B.~M. Wiernik, and P.~Rosen.
\newblock Automatic scatterplot design optimization for clustering identification.
\newblock {\em IEEE Trans. Vis. Comput. Graph.}, 2022. \href{https://doi.org/10.1109/tvcg.2022.3189883}
{doi: {{%
10\hspace{.1pt}\discretionary{.}{%
}{.}\hspace{.4pt}1109\discretionary{/}{%
}{/}tvcg\hspace{.1pt}\discretionary{.}{%
}{.}\hspace{.4pt}2022\hspace{.1pt}\discretionary{.}{%
}{.}\hspace{.4pt}3189883}}}


\bibitem{quadri2020modeling}
G.~J. Quadri and P.~Rosen.
\newblock Modeling the influence of visual density on cluster perception in scatterplots using topology.
\newblock {\em IEEE Trans. Vis. Comput. Graph.}, 27(2):1829--1839, 2020. \href{https://doi.org/10.1109/tvcg.2020.3030365}
{doi: {{%
10\hspace{.1pt}\discretionary{.}{%
}{.}\hspace{.4pt}1109\discretionary{/}{%
}{/}tvcg\hspace{.1pt}\discretionary{.}{%
}{.}\hspace{.4pt}2020\hspace{.1pt}\discretionary{.}{%
}{.}\hspace{.4pt}3030365}}}


\bibitem{quadri2021survey}
G.~J. Quadri and P.~Rosen.
\newblock A survey of perception-based visualization studies by task.
\newblock {\em IEEE Trans. Vis. Comput. Graph.}, 28(12), 2021. \href{https://doi.org/10.1109/tvcg.2021.3098240}
{doi: {{%
10\hspace{.1pt}\discretionary{.}{%
}{.}\hspace{.4pt}1109\discretionary{/}{%
}{/}tvcg\hspace{.1pt}\discretionary{.}{%
}{.}\hspace{.4pt}2021\hspace{.1pt}\discretionary{.}{%
}{.}\hspace{.4pt}3098240}}}


\bibitem{quadri2024do}
G.~J. Quadri, A.~Z. Wang, Z.~Wang, J.~Adorno, P.~Rosen, and D.~A. Szafir.
\newblock Do you see what i see? a qualitative study eliciting high-level visualization comprehension.
\newblock In {\em ACM CHI}, pp. 1--26, 2024. \href{https://doi.org/10.1145/3613904.3642813}
{doi: {{%
10\hspace{.1pt}\discretionary{.}{%
}{.}\hspace{.4pt}1145\discretionary{/}{%
}{/}3613904\hspace{.1pt}\discretionary{.}{%
}{.}\hspace{.4pt}3642813}}}


\bibitem{quadri2021constructing}
G.~J. A.~R. Quadri.
\newblock {\em Constructing Frameworks for Task-Optimized Visualizations}.
\newblock PhD thesis, University of South Florida, 2021.

\bibitem{R}
{R Core Team}.
\newblock {\em R: A Language and Environment for Statistical Computing}.
\newblock R Foundation for Statistical Computing, Vienna, Austria, 2021.

\bibitem{rensink2010perception}
R.~Rensink and G.~Baldridge.
\newblock The perception of correlation in scatterplots.
\newblock {\em Comput. Graph. Forum}, 29(3):1203--1210, 2010. \href{https://doi.org/10.1111/j.1467-8659.2009.01694.x}
{doi: {{%
10\hspace{.1pt}\discretionary{.}{%
}{.}\hspace{.4pt}1111\discretionary{/}{%
}{/}j\hspace{.1pt}\discretionary{.}{%
}{.}\hspace{.4pt}1467\discretionary{%
}{-}{-}8659\hspace{.1pt}\discretionary{.}{%
}{.}\hspace{.4pt}2009\hspace{.1pt}\discretionary{.}{%
}{.}\hspace{.4pt}01694\hspace{.1pt}\discretionary{.}{%
}{.}\hspace{.4pt}x}}}


\bibitem{rensink2018information}
R.~A. Rensink.
\newblock Information visualization and the study of visual perception.
\newblock {\em Journal of Vision}, 18(10):1350--1350, 2018. \href{https://doi.org/10.1167/18.10.1350}
{doi: {{%
10\hspace{.1pt}\discretionary{.}{%
}{.}\hspace{.4pt}1167\discretionary{/}{%
}{/}18\hspace{.1pt}\discretionary{.}{%
}{.}\hspace{.4pt}10\hspace{.1pt}\discretionary{.}{%
}{.}\hspace{.4pt}1350}}}


\bibitem{sarikaya2018scatterplots}
A.~Sarikaya and M.~Gleicher.
\newblock Scatterplots: Tasks, data, and designs.
\newblock {\em IEEE Trans. Vis. Comput. Graph.}, 24(1):402--412, 2018.

\bibitem{schloss2018mapping}
K.~B. Schloss, C.~C. Gramazio, A.~T. Silverman, M.~L. Parker, and A.~S. Wang.
\newblock Mapping color to meaning in colormap data visualizations.
\newblock {\em IEEE Trans. Vis. Comput. Graph.}, 25(1):810--819, 2018. \href{https://doi.org/10.1109/TVCG.2018.2865147}
{doi: {{%
10\hspace{.1pt}\discretionary{.}{%
}{.}\hspace{.4pt}1109\discretionary{/}{%
}{/}TVCG\hspace{.1pt}\discretionary{.}{%
}{.}\hspace{.4pt}2018\hspace{.1pt}\discretionary{.}{%
}{.}\hspace{.4pt}2865147}}}


\bibitem{schloss2011aesthetic}
K.~B. Schloss and S.~E. Palmer.
\newblock Aesthetic response to color combinations: preference, harmony, and similarity.
\newblock {\em Atten. Percept. Psychophys.}, 73:551--571, 2011. \href{https://doi.org/10.3758/s13414-010-0027-0}
{doi: {{%
10\hspace{.1pt}\discretionary{.}{%
}{.}\hspace{.4pt}3758\discretionary{/}{%
}{/}s13414\discretionary{%
}{-}{-}010\discretionary{%
}{-}{-}0027\discretionary{%
}{-}{-}0}}}


\bibitem{setlur2016linguistic}
V.~Setlur and M.~Stone.
\newblock A linguistic approach to categorical color assignment for data visualization.
\newblock {\em IEEE Trans. Vis. Comput. Graph.}, 22, 2016. \href{https://doi.org/10.1109/TVCG.2015.2467471}
{doi: {{%
10\hspace{.1pt}\discretionary{.}{%
}{.}\hspace{.4pt}1109\discretionary{/}{%
}{/}TVCG\hspace{.1pt}\discretionary{.}{%
}{.}\hspace{.4pt}2015\hspace{.1pt}\discretionary{.}{%
}{.}\hspace{.4pt}2467471}}}


\bibitem{shaw1998data}
C.~D. Shaw, D.~S. Ebert, J.~M. Kukla, A.~Zwa, I.~Soboroff, and D.~A. Roberts.
\newblock Data visualization using automatic perceptually motivated shapes.
\newblock In {\em Visual Data Exploration and Analysis V}, vol. 3298, pp. 208--213. SPIE, 1998. \href{https://doi.org/10.1117/12.309543}
{doi: {{%
10\hspace{.1pt}\discretionary{.}{%
}{.}\hspace{.4pt}1117\discretionary{/}{%
}{/}12\hspace{.1pt}\discretionary{.}{%
}{.}\hspace{.4pt}309543}}}


\bibitem{smart2019measuring}
S.~Smart and D.~A. Szafir.
\newblock Measuring the separability of shape, size, and color in scatterplots.
\newblock In {\em Proc. ACM Hum. Factors Comput. Syst. (CHI)}, p. 669, 2019. \href{https://doi.org/10.1145/3290605.3300899}
{doi: {{%
10\hspace{.1pt}\discretionary{.}{%
}{.}\hspace{.4pt}1145\discretionary{/}{%
}{/}3290605\hspace{.1pt}\discretionary{.}{%
}{.}\hspace{.4pt}3300899}}}


\bibitem{stone2006choosing}
M.~Stone.
\newblock Choosing colors for data visualization.
\newblock {\em Business Intelligence Network}, 2, 2006.

\bibitem{szafir2018modeling}
D.~A. Szafir.
\newblock Modeling color difference for visualization design.
\newblock {\em IEEE Trans. Vis. Comput. Graph.}, 24(1):392--401, 2018. \href{https://doi.org/10.1109/TVCG.2017.2744359}
{doi: {{%
10\hspace{.1pt}\discretionary{.}{%
}{.}\hspace{.4pt}1109\discretionary{/}{%
}{/}TVCG\hspace{.1pt}\discretionary{.}{%
}{.}\hspace{.4pt}2017\hspace{.1pt}\discretionary{.}{%
}{.}\hspace{.4pt}2744359}}}


\bibitem{szafir2023visualization}
D.~A. Szafir, R.~Borgo, M.~Chen, D.~J. Edwards, B.~Fisher, and L.~Padilla.
\newblock {\em Visualization Psychology}.
\newblock Springer Nature, 2023.

\bibitem{szafir2016four}
D.~A. Szafir, S.~Haroz, M.~Gleicher, and S.~Franconeri.
\newblock Four types of ensemble coding in data visualizations.
\newblock {\em Journal of Vision}, 16(5):11--11, 2016. \href{https://doi.org/10.1167/16.5.11}
{doi: {{%
10\hspace{.1pt}\discretionary{.}{%
}{.}\hspace{.4pt}1167\discretionary{/}{%
}{/}16\hspace{.1pt}\discretionary{.}{%
}{.}\hspace{.4pt}5\hspace{.1pt}\discretionary{.}{%
}{.}\hspace{.4pt}11}}}


\bibitem{tableau}
Tableau.
\newblock Tableau, 2022.

\bibitem{tseng2023evaluating}
C.~Tseng, G.~J. Quadri, Z.~Wang, and D.~A. Szafir.
\newblock Measuring categorical perception in color-coded scatterplots.
\newblock In {\em Proc. ACM Hum. Factors Comput. Syst. (CHI)}, 2023. \href{https://doi.org/10.1145/3544548.3581416}
{doi: {{%
10\hspace{.1pt}\discretionary{.}{%
}{.}\hspace{.4pt}1145\discretionary{/}{%
}{/}3544548\hspace{.1pt}\discretionary{.}{%
}{.}\hspace{.4pt}3581416}}}


\bibitem{tseng2024revisiting}
C.~Tseng, A.~Z. Wang, G.~J. Quadri, and D.~A. Szafir.
\newblock Revisiting categorical color perception in scatterplots: Sequential, diverging, and categorical palettes.
\newblock In {\em Proceedings of the 26th EG/VGTC Conference on Visualization (EuroVis)}, 2024. \href{https://doi.org/10.2312/evs.20241073}
{doi: {{%
10\hspace{.1pt}\discretionary{.}{%
}{.}\hspace{.4pt}2312\discretionary{/}{%
}{/}evs\hspace{.1pt}\discretionary{.}{%
}{.}\hspace{.4pt}20241073}}}


\bibitem{urribarri2017prediction}
D.~K. Urribarri and S.~M. Castro.
\newblock Prediction of data visibility in two-dimensional scatterplots.
\newblock {\em Information Visualization}, 16(2):113--125, 2017. \href{https://doi.org/10.1177/1473871616638892}
{doi: {{%
10\hspace{.1pt}\discretionary{.}{%
}{.}\hspace{.4pt}1177\discretionary{/}{%
}{/}1473871616638892}}}


\bibitem{wang2019improving}
Y.~Wang, Z.~Wang, T.~Liu, M.~Correll, Z.~Cheng, O.~Deussen, and M.~Sedlmair.
\newblock Improving the robustness of scagnostics.
\newblock {\em IEEE Trans. Vis. Comput. Graph.}, 26(1):759--769, 2019. \href{https://doi.org/10.1109/TVCG.2019.2934796}
{doi: {{%
10\hspace{.1pt}\discretionary{.}{%
}{.}\hspace{.4pt}1109\discretionary{/}{%
}{/}TVCG\hspace{.1pt}\discretionary{.}{%
}{.}\hspace{.4pt}2019\hspace{.1pt}\discretionary{.}{%
}{.}\hspace{.4pt}2934796}}}


\bibitem{ware2009quantitative}
C.~Ware.
\newblock Quantitative texton sequences for legible bivariate maps.
\newblock {\em IEEE Transactions on Visualization and Computer Graphics}, 15(6):1523--1530, 2009. \href{https://doi.org/10.1109/TVCG.2009.175}
{doi: {{%
10\hspace{.1pt}\discretionary{.}{%
}{.}\hspace{.4pt}1109\discretionary{/}{%
}{/}TVCG\hspace{.1pt}\discretionary{.}{%
}{.}\hspace{.4pt}2009\hspace{.1pt}\discretionary{.}{%
}{.}\hspace{.4pt}175}}}


\bibitem{ware2012information}
C.~Ware.
\newblock {\em Information visualization: perception for design}.
\newblock Elsevier, 2012.

\bibitem{wilkinson2005graph}
L.~Wilkinson, A.~Anand, and R.~Grossman.
\newblock Graph-theoretic scagnostics.
\newblock In {\em IEEE Symp. Info. Vis. (INFOVIS)}, pp. 157--164. IEEE, Oct 2005. \href{https://doi.org/10.1109/INFVIS.2005.1532142}
{doi: {{%
10\hspace{.1pt}\discretionary{.}{%
}{.}\hspace{.4pt}1109\discretionary{/}{%
}{/}INFVIS\hspace{.1pt}\discretionary{.}{%
}{.}\hspace{.4pt}2005\hspace{.1pt}\discretionary{.}{%
}{.}\hspace{.4pt}1532142}}}


\end{thebibliography}
